\documentclass[11pt]{iopart}
\usepackage{graphicx}
\usepackage{subfigure}
\usepackage{morefloats}
\usepackage{color}

\begin{document}

\title{Characterization of 3D filament dynamics in a MAST SOL flux tube geometry}

\author{N. R. Walkden$^{1,2}$, B. D. Dudson$^{2}$ and  G. Fishpool$^{1}$
        \\ \small{$^{1}$EURATOM/CCFE Fusion Association, Culham Science Centre, Abingdon, OX14 3DB, UK} 
        \\ \small{$^{2}$York Plasma Institute, Department of Physics, University of York, Heslington, York, YO10 5DD, UK} 
        \\ Email: \texttt{nrw504@york.ac.uk} }
\date{}
\begin{abstract}
Non-linear simulations of filament propagation in a realistic MAST SOL flux tube geometry using the BOUT++ fluid modelling framework show an isolation of the dynamics of the filament in the divertor region from the midplane region due to three features of the magnetic geometry; the variation of magnetic curvature along the field line, the expansion of the flux tube and strong magnetic shear. Of the three effects, the latter two lead to a midplane ballooning feature of the filament, whilst the former leads to a ballooning around the X-points. In simulations containing all three effects the filament is observed to balloon at the midplane, suggesting that the role of curvature variation is sub-dominant to the flux expansion and magnetic shear. The magnitudes of these effects are all strongest near the X-point which leads to the formation of parallel density gradients. The filaments simulated, which represent filaments in MAST, are identified as resistive ballooning, meaning that their motion is inertially limited, not sheath limited. Parallel density gradients can drive the filament towards a Boltzmann response when the collisionalityof the plasma is low. The results here show that the formation of parallel density gradients is a natural and inevitable consequence of a realistic magnetic geometry and therefore the transition to the Boltzmann response is a consequence of the use of realistic magnetic geometry and does not require initializing specifically varying background profiles as in slab simulations. The filaments studied here are stable to the linear resistive drift wave instability but are subject to the non-linear effects associated with the Boltzmann response, particularly Boltzmann spinning. The Boltzmann response causes the filament to self-organise and spin on an axis. In later stages of its evolution a non-linear turbulent state develops where the vorticity evolves into a turbulent eddy field on the same length scale as the parallel current. The transition from interchange motion to the Boltzmann response occurs with increasing temperature through a decrease in collisionality. This is confirmed by measuring the correlation between density and potential perturbations within the filament, which is low in the anti-symmetric state associated with the interchange mechanism, but high in the Boltzmann regime. In the Boltzmann regime net radial transport is drastically reduced whilst a small net toroidal transport is observed. This suggests that only a subset of filaments, those driven by the interchange mechanism at the separatrix, can propagate into the far SOL. Filaments in the Boltzmann regime will be confined to the near separatrix region and quickly disperse. It is plausible that filaments in both regimes can contribute to the SOL transport observed in experiment; the former by propagating the filament into the far SOL and the latter by dispersion of the density within the filament. 
\end{abstract}

\section{Introduction}
\begin{figure}[htp]
\includegraphics[width=6cm]{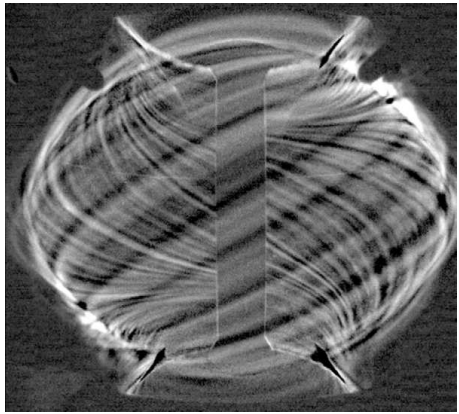}
\centering
\caption{Visible light image of filaments in Ohmic L-mode on MAST. The fast varying component of light has been digitally enhanced. Reprinted from \cite{DudsonPPCF2008} with permission from B. D. Dudson.}
\label{MAST_photo}
\end{figure}

\begin{figure}[ht]
\includegraphics[width=5cm]{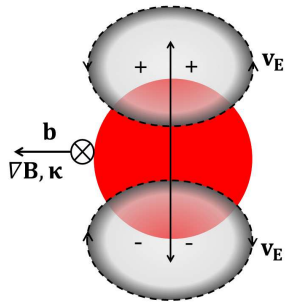}
\centering
\caption{Schematic example of the physics of blob polarization. Curvature and $\nabla B$ drifts polarize the blob which leads to a radial $\textbf{E}\times\textbf{B}$ drift formed of two counter rotating vortices.}
\label{Blob_schematic1}
\end{figure}

\begin{figure}[ht]
\includegraphics[width=5cm]{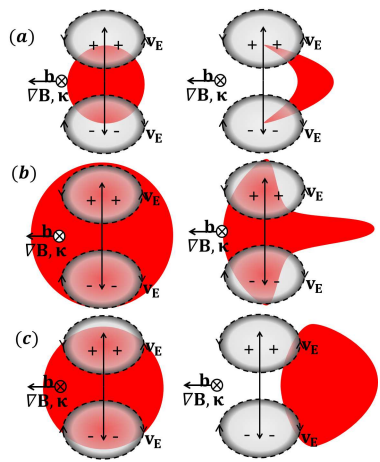}
\centering
\caption{Schematic diagram of blob structure from: a. mushrooming motion, b. fingering motion and c. coherent propagation}
\label{blob_schematic2}
\end{figure}

Filaments are field aligned plasma structures that propagate in the scrape off layer (SOL) region of magnetically confined plasmas. They have been observed on a number of tokamaks ~\cite{BoedoPoP2001,KirkPPCF2006,DudsonPPCF2005,DudsonPPCF2008,CarrerasPoP2001,EndlerNF95,ZwebenPoP2002} and other magnetic confinement devices \cite{FasoliPoP2006,AntarPRL2001,YamadaNATURE2008}. Filaments carry a significant number of particles into the SOL and can play a dominant role in determining L-mode and inter-ELM H-mode \cite{AyedPPCF2009} SOL properties. Modelling of the SOL is often used in the design of future magnetic confinement devices; however an incomplete understanding of non-diffusive plasma transport limits the accuracy of such predictions. Filament propagation in the SOL is a competition between perpendicular drift motion and parallel streaming \cite{KrasheninnikovPhysLet2001}; the former transporting the filament radially and the latter draining the filament to the divertor target surface. If the filament can propagate radially fast enough it may contact first wall material surfaces causing damage and erosion. Alternatively if filaments propagate slowly they may drain to localised spots on the target surface, leading to undesirable hot spots. The challenge of predicting particle loading on material surfaces therefore necessitates an understanding of filament dynamics. 
\\ \\Filaments are transient phenomena with a lifetime not exceeding a few $100$s of $\mu s$ on the Mega Amp Spherical Tokamak (MAST)\cite{DudsonPPCF2008}. They are highly localised in the drift plane. Generally they  are observed to extend from at least X-point to X-point on MAST. Figure \ref{MAST_photo} shows these features in a fast camera image of a MAST L-mode plasma with digital enhancement of the fast varying component of the light. 
The disparity between parallel and perpendicular dynamics in the filament, arising from the fast streaming of particles along the magnetic field, leads to their consideration as quasi-2D objects in the drift plane with closure schemes employed to account for dynamics in the third (parallel) dimension \cite{KrasheninnikovPhysLet2001}. The most common closure is the sheath-limited scheme where the parallel current is equated to the sheath current \cite{NedospasovNF85} at the divertor plate. A similar scheme will be used here. Other closure schemes consider, for example, an enhanced polarization current by cross-field resistivity near plasma X-points, or Alfven-wave generation in high $\beta$ plasmas \cite{KrasheninnikovCzechJourn2005,YuPoP2006}. These 2D objects are often termed '\emph{blobs}'. Blob dynamics has received increased attention in recent years \cite{KrasheninnikovCzechJourn2005,YuPoP2006,GarciaPoP2006,BianPoP2003,MyraPoP2004,YuPoP2003,JovanovicPoP2008,RozhanskyPPCF2008,MadsenPoP2011,KrasheninnikovPoP2003,D'IppolitoPoP2004,D'IppolitoContribPlasPhys2004,RussellPRL2004,BodiPoP2008}. For a review of the subject see \cite{D'IppolitoReview}.
\\ \\The basis of blob physics is described in figure \ref{Blob_schematic1}. 
In a tokamak, curvature and $\nabla B$ forces polarize the blob by inducing charge dependant drift motion on the ions and electrons. The polarization of the blob gives rise to an $\textbf{E}\times\textbf{B}$ velocity field which takes the form of a pair of counter rotating vortices, advecting the blob outwards. This configuration often leads to the classic '\emph{mushroom}' shape observed in many simulations, \cite{GarciaPoP2006,YuPoP2003,D'IppolitoPoP2004,YuPoP2006} for example, and in experiment \cite{KatzPRL2008}. The proceeding dynamics of the blob is then determined by the means of charge dissipation in the model. In the 2D picture presented above charge is dissipated by sheath currents. The subsequent perpendicular motion depends on the size of the blob; a small blob will mushroom since the induced vortices are partly external to the blob itself, whilst a large blob will form a fingering structure (reminiscent of the familiar Rayleigh Taylor instability) since the vortices are entirely internal \cite{KrasheninnikovPhysLet2001,YuPoP2006,YuPoP2003}. At some critical blob size these two effects balance and the blob may propagate coherently for many times its width \cite{KrasheninnikovPhysLet2001}. These three forms of motion are detailed schematically in figure \ref{blob_schematic2}. 
If the filament is not in contact with the divertor target then cross-field resistivity is the dominant charge dissipation mechanism. In the former case the filament exhibits the mushrooming characteristic, with smaller scale structure appearing depending on the viscosity and resistivity of the plasma \cite{GarciaPoP2006} due to Kelvin-Helmholtz instabilities. Recently accounting for the full 3D nature of filaments has been shown to lead to a significant departure of the blob dynamics from 2D theory \cite{AngusPRL2012,AngusPoP2012,JovanovicPoP2008}. When the linear growth rate of the resistive drift-wave instability is larger than the interchange growth rate on the blob front (with respect to propagation direction) the blob becomes unstable to resistive drift waves \cite{AngusPRL2012}. The blob becomes much more diffuse and less coherent than its 2D counterpart. Drift-waves require a dissipation mechanism to become unstable. This can either be through collisions or wave-particle interactions. In this case the dissipation comes from resistivity which increases with collisionality. When the collisionality is low resistive dissipation is reduced and drift-waves become stable. Instead the filament is driven towards a Boltzmann response which causes the blob to spin about its center \cite{AngusPoP2012}, suppressing the mushrooming motion. Such spinning can lead to Kelvin-Helmholtz instabilities \cite{D'IppolitoPoP2004} which break the blob up, or simply reduced the radial advection by dissipating the polarized charge. 
\\ \\A significant body of work exists which describes the basic motion of filaments, however only a limited subset of this work considers the effects of a realistic magnetic geometry. The magnetic geometry in the SOL will potentially influence the motion of the filament. The drive for interchange motion arises due to the magnetic curvature which can vary dramatically along the length of a filament. Consequently the advection velocity of the filament may vary along its length, leading to the formation of parallel density gradients which have been shown to affect the cross-field dynamics of the filament. The question then arises as to what significance this gives to 3D effects in a realistic magnetic geometry and how these effects might manifest in realistic plasmas. In this paper this question will be addressed by conducting full 3D simulations of filaments in the BOUT++ fluid modelling framework using a drift-reduced Braginskii model\cite{BOUT++}. The simulation domain is a flux tube based on an EFIT \cite{EFIT} equilibrium reconstruction of a MAST ohmic L-mode plasma, shot 14220. The model is described in section 2. Section 3 describes the MAST flux tube simulation domain. Section 4 presents an investigation of the effects of the magnetic geometry on the filament dynamics whilst section 5 investigates the role of 3D effects in the cross-field dynamics of the filament. Finally section 6 concludes.

\section{Governing equations}

\begin{figure}[ht]
\includegraphics[width=7cm]{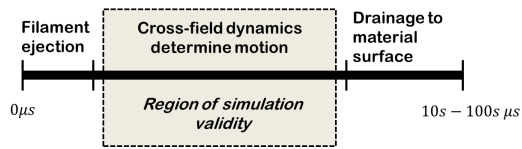}
\centering
\caption{Time line of a filament from its ejection into the SOL until it drains to a material surface. The shaded region indicates roughly the region of validity for the model used to simulate filaments herein.}
\label{filament_lifetime}
\end{figure}

The governing equations for the model of filament dynamics presented here are derived from a drift ordered reduction of the Braginskii equations \cite{Braginskii}. Following \cite{SimakovPoP2003,SimakovPoP2004} and assuming an isothermal, electrostatic plasma, neglecting electron inertia and assuming cold ions the governing equations in SI units are the density equation;
\begin{equation}
\frac{dn}{dt} = 2\rho_{s}c_{s}\xi\cdot\left(\nabla n - n\nabla\phi\right) + \frac{1}{e}\nabla_{||}J_{||} - n\nabla_{||}u_{||}
\end{equation}
the vorticity equation;
\begin{equation}
\rho_{s}^{2}\nabla\cdot \left(n\frac{d\nabla_{\perp}\phi}{dt}\right) = 2\rho_{s}c_{s}\xi\cdot\nabla n + \frac{1}{e}\nabla_{||}J_{||}
\end{equation}
the parallel momentum equation;
\begin{equation}
\frac{du_{||}}{dt} = -\frac{c_{s}^{2}}{n}\nabla_{||}n
\end{equation}
where 
\[\frac{d}{dt} \equiv \frac{\partial}{\partial t} + c_{s}\rho_{s}\textbf{b}\times\nabla\phi \cdot\nabla \]
and parallel Ohm's law;
\begin{equation}
J_{||} = \frac{\sigma_{||}T_{e}}{ne}\left(\nabla_{||}n - n\nabla_{||}\phi\right)
\end{equation}
In deriving these equation the assumption
\begin{equation}
\nabla\times\left(\frac{\textbf{b}}{B}\right) \approx  \frac{2}{B}\xi = \frac{2}{B}\textbf{b}\times\kappa
\end{equation}
is made, where $\kappa$ is the magnetic curvature vector and $\textbf{b}$ is the magnetic field tangency vector. $\xi$ (a vector), which is defined by (5), is the polarization vector and defines the strength and direction of polarization due to curvature forcing. $n$ is the plasma density and $\phi$ is the normalized electrostatic plasma potential given by 
\begin{equation}
\phi = \frac{e\Phi}{T_{e}}
\end{equation}
where $\Phi$ is the plasma potential. $J_{||}$ is the parallel current density and $u_{||}$ is the parallel ion velocity. $\rho_{s}$ and $c_{s}$ are the Bohm gyro-radius and sound speed given by 
\begin{equation}
\begin{array}{c c c}
\rho_{s} = c_{s}/\Omega_{i}, & c_{s}^{2} = T_{e}/m_{i}, & \Omega_{i} = eB/m_{i}
\end{array}
\end{equation}
$T_{e}$ is the (isothermal) electron temperature, $B$ is the magnetic field strength and $m_{i}$ is the ion mass. Finally $\sigma_{||}$ is the collisional parallel conductivity \cite{Braginskii} given by 
\begin{equation}
\sigma_{||} = 1.96 \frac{ne^{2}\tau_{e}}{m_{e}}
\end{equation}
where the electron collision time, $\tau_{e}$ is 
\begin{equation}
\tau_{e} = \frac{3\sqrt{m_{e}}T_{e}^{3/2}}{4\sqrt{2\pi}n\lambda e^{4}} = 3.44\times 10^{11}\frac{T_{e}\left(eV\right)^{3/2}}{n\lambda}
\end{equation}
where $\lambda \sim 10 $ is the Coulomb logarithm. 
\\ \\These equations omit major aspects of physics in the SOL yet retain the core physics controlling filament propagation, making them ideal for a theoretical study. Neglecting electromagnetic effects limits the model to the case of low $\beta$ plasmas. This is a fragile assumption for MAST which is a high $\beta$ machine, even near the plasma edge. Furthermore ions have been shown to exhibit comparable and often greater temperatures than electrons \cite{KirkPPCF2004,KocanPPCF2010,KocanJNucMat2011,TamainJNucMat2011,ElmorePPCF2012,ElmoreFusEngDes2013,AllanJNucMat2013} making the cold ion assumption suspect. Hot ions have been shown to affect filaments \cite{JovanovicPoP2008} and will be studied in 3D in a future paper. For this work the purpose is to study the basic mechanisms underlying filament motion, which is predominantly controlled by electron dynamics, so hot ions have been neglected in this study. In the non-isothermal case hot ions can cause filaments to be ejected at an angle to the radial direction \cite{BisaiPoP2012}. It is worth noting however, that hot ions also tend to reinforce the interchange mechanism in non-thermalized blobs \cite{BisaiPoP2012}. In the interchange mechanism the main role of hot ions is to increase the pressure of the plasma which enhances the drive for the interchange motion, but does not change the motion itself. When 3D effects are important hot ions can affect the stability of resistive drift waves \cite{AngusDriftWave2012}, though the effect is not significant and certainly does not alter the basic mechanism behind the resistive drift wave. The inclusion of hot ions is therefore not essential to modelling filament motion and consequently has been neglected here.  The effect of a neutral particle species is also not included. Whilst this is a common assumption in blob modelling, it is important to recognise that the SOL has a large population of neutral particles. Despite their limitations, equations (1-4) allow the basic mechanisms controlling advective motion of coherent plasma structures to be probed. Similar systems have been used in 2D to study the turbulent region at the plasma edge which is known to eject filaments \cite{MilitelloPPCF2013,GarciaPPCF2006} and produce excellent agreement with experiment. This suggests that simple models of this form at least capture the basic physical mechanisms controlling the ejection and evolution of filaments. Thomson scattering measurements on MAST \cite{KirkPPCF2006} show that the density perturbation of a filament far outweighs the temperature perturbation, which motivates the isothermal approximation used within this model. To further reduce the system the parallel ion terms are neglected. This is valid on time-scales 
\begin{equation}
\tau < L_{||}/2c_{s} \sim 10^{-4}s
\end{equation}
where $L_{||}/2c_{s}$ is the time taken for a parallel sound wave to propagate from the centre of the filament to the target. On time-scales shorter than this the filament is dominated by advective cross-field motion. This represents the early to intermediate stages of filament evolution in the SOL, as indicated in figure \ref{filament_lifetime}. 
The $\textbf{E}\times\textbf{B}$ drift is considered incompressible allowing neglect of the $n\nabla\phi$ term in the density equation. The Boussinesque approximation is also invoked such that 
\begin{equation}
\nabla\cdot\left(n\frac{d\nabla_{\perp}\phi}{dt}\right)  \approx n\frac{d}{dt}\nabla_{\perp}^{2}\phi
\end{equation}
The Boussinesque approximation is strictly only valid for small perturbations. Filaments with $\delta n / n \sim 1$ certainly do not satisfy this condition and it has been shown that solving the full expression can lead to augmentations of the blobs cross-field structure \cite{YuPoP2006}. Solving the full expression is numerically demanding and, since the relaxation of the Boussinesque approximation does not alter the basic dynamic mechanisms within the system, it has been made here.
\\ \\Applying these further reductions gives the system
\begin{equation}
\frac{dn}{dt} = 2\rho_{s}c_{s}\xi\cdot\nabla n + \frac{1}{e}\nabla_{||}J_{||} 
\end{equation}
\begin{equation}
\rho_{s}^{2}n\frac{d\nabla^{2}_{\perp}\phi}{dt} = 2\rho_{s}c_{s}\xi\cdot\nabla n + \frac{1}{e}\nabla_{||}J_{||} 
\end{equation}
and equation (4) for $J_{||}$. This is the system employed in the 3D studies by Angus \emph{et. al}\cite{AngusPRL2012,AngusPoP2012}.
The system defined by equations (12), (13) and (4) supports the linear interchange instability and the linear resistive drift wave instability \cite{AngusPRL2012,AngusPoP2012}. The interchange instability is destabilized by perpendicular gradients in pressure, or (since the model is isothermal) density gradients, such as those at the blob front. The resistive drift wave is destabilized by collisional dissipation in the form of resistivity. Angus has shown that the relative importance of unstable drift waves in blob propagation is determined by blob size, with larger blobs being more stable to resistive drift waves and therefore holding their shape. In this study filaments simulations have been conducted with resistive drift-waves initially stable by the condition of Angus \emph{et.al} \cite{AngusPRL2012}. This allows a clear investigation into the role of Boltzmann response at low collisionality in filament dynamics without the added complication of unstable linear resistive drift-waves as collisionality increases.

\section{Simulation geometry}

\begin{figure}[ht!]
  \begin{center}
    \subfigure[MAST flux surfaces]{
      \label{Metric_quantities:MAST_surfaces}
      \includegraphics[width=0.2\textwidth]{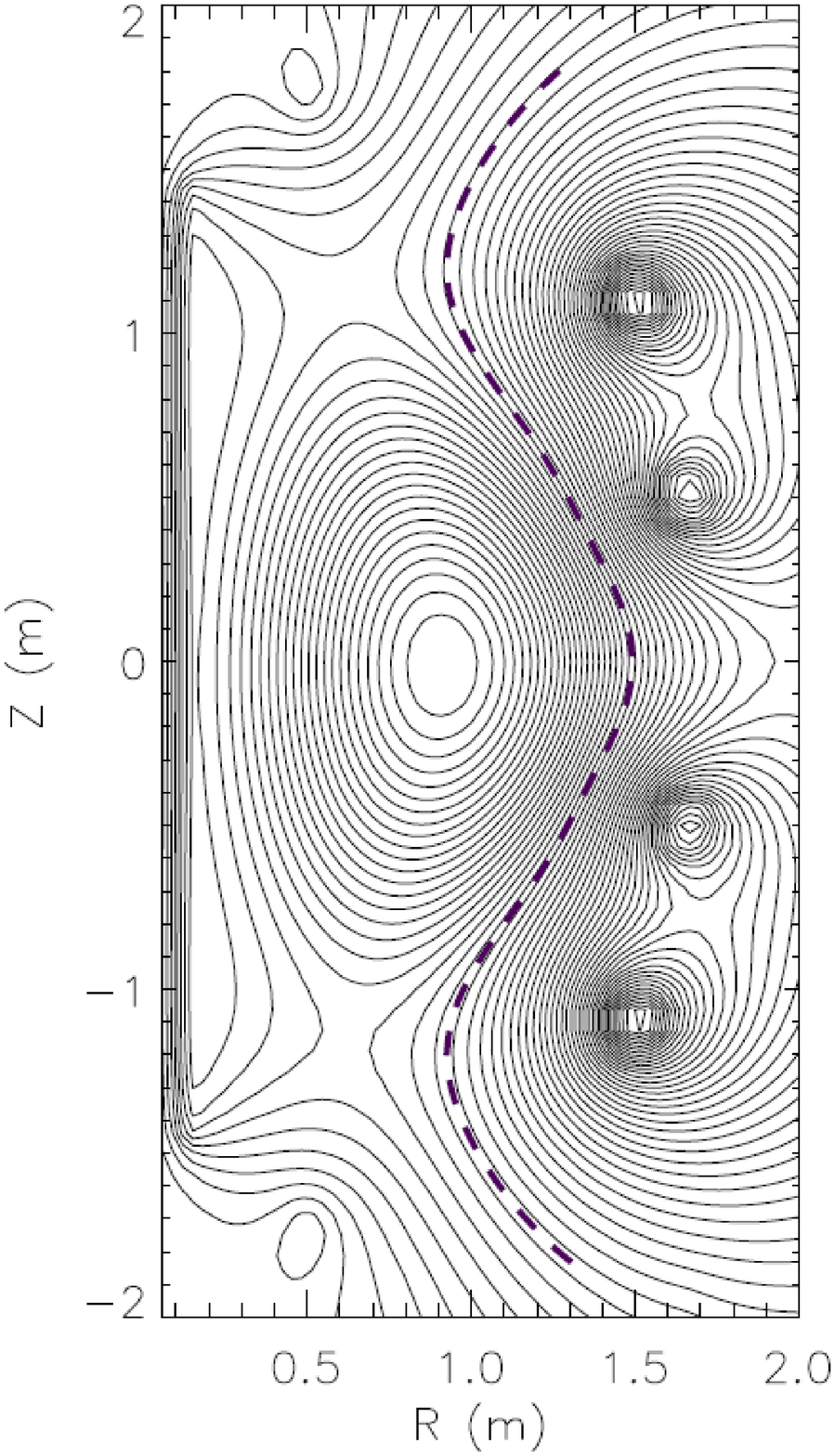}
      }
    \subfigure[Magnetic field]{
      \includegraphics[width=0.3\textwidth]{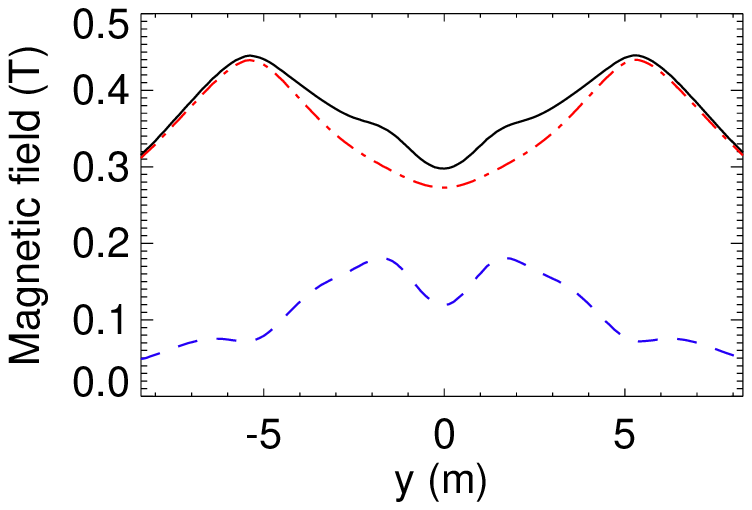}
      \label{Metric_quantities:B-field}
      }
    \subfigure[Shear (broken) and Integrated shear (solid)]{
      \label{Metric_quantities:Shear}
      \includegraphics[width=0.3\textwidth]{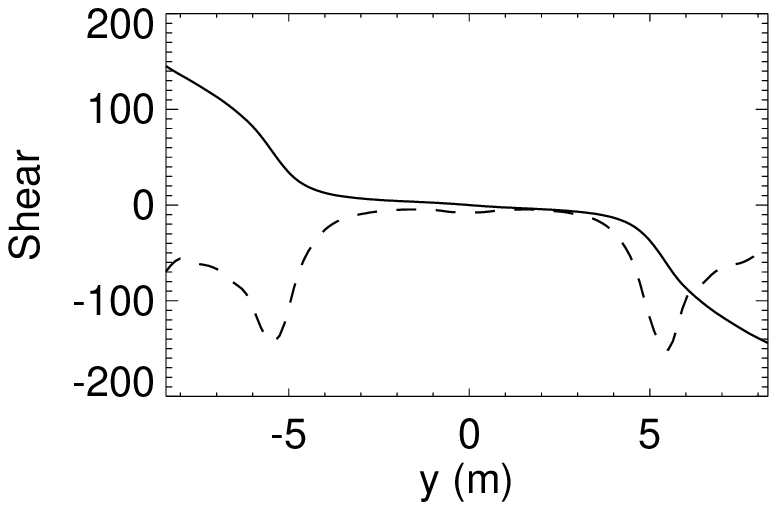}
      }
    \end{center}
  \caption{Equilibrium components to the coordinate system metric, consisting Total (solid, black), poloidal (dashed, blue) and toroidal (dot-dash, red) components of magnetic field (b)  and magnetic shear (c) derived on a field line in the MAST SOL (a)}
\label{Metric_quantities}
\end{figure}

\begin{figure}[ht!]
\includegraphics[width=0.4\textwidth]{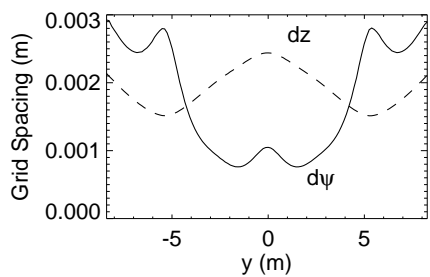}
\centering
\caption{Real space grid spacing along the length of the flux tube. To achieve comparable grid spacings in $x$ and $z$ the $x$ domain is scaled by $7$.}
\label{grid_spacing}
\end{figure}

\begin{figure}[ht!]
\begin{center}
  \subfigure[$z$ component]{
    \includegraphics[width=0.3\textwidth]{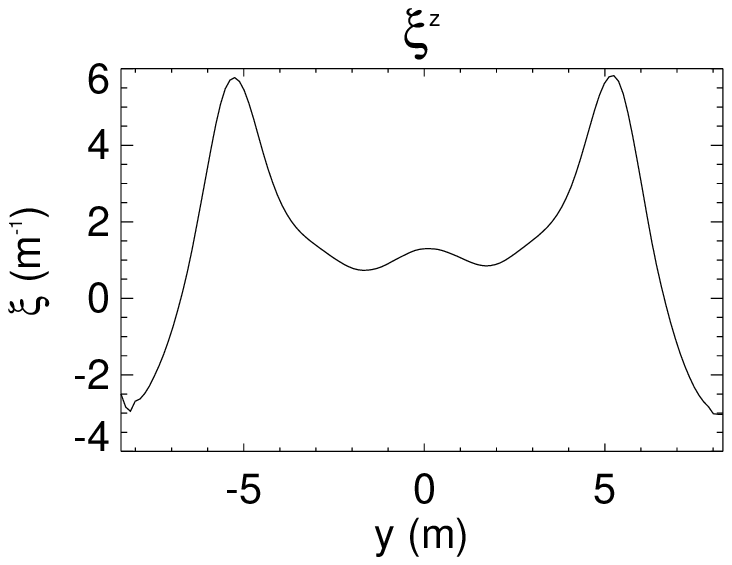}
    \label{curvature:z}
    }
  \subfigure[$x$ component]{
    \includegraphics[width=0.3\textwidth]{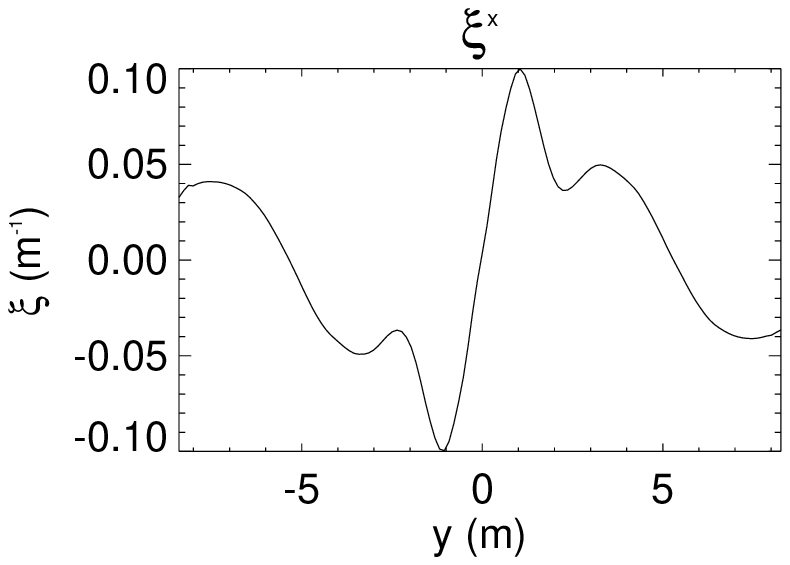}
    \label{curvature:x}
    }
  \subfigure[$\psi$ (driving) component]{
    \includegraphics[width=0.3\textwidth]{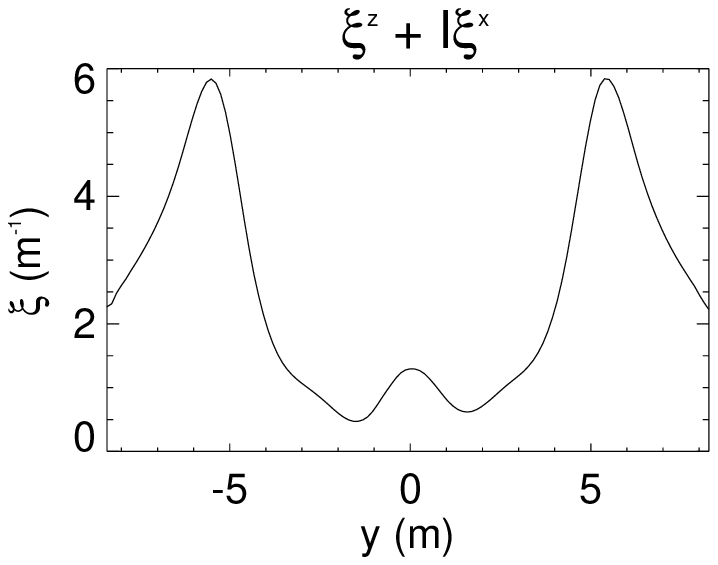}
    \label{curvature:psi}
    }
\end{center}
  \caption{Contravariant components of the polarization vector $\xi = \textbf{b}\times\kappa$ where $\kappa$ is the magnetic curvature vector.}
  \label{curvature}
\end{figure}

The field aligned nature of SOL filaments makes a field aligned coordinate system ideal for their study. The field aligned system, $\left(x,y,z\right)$ is defined on the LFS by \cite{XuBOUT}
\begin{equation}
\begin{array}{c}
x = \psi_{N} - \psi_{N}^{(0)} \\
y = \theta \\
z = \zeta - \int_{\theta_{0}}^{\theta}\nu\left(\psi_{N},\theta ' \right)d\theta '
\end{array}
\end{equation}
where $\psi_{N}$ is the normalized poloidal flux and $\psi_{N}^{(0)}$ defines the center of the flux tube, here taken as $\psi_{N}^{(0)} = 1.15$. $\theta$ is the poloidal angle, $\zeta$ is the toroidal angle and $\nu$ is the local field line pitch. The modification of the toroidal angle in $z$ ensures that $y$ indeed follows the magnetic field line whilst $z$ retains toroidal periodicity, allowing Fourier techniques to be used in the $z$ coordinate. The covariant basis vectors (vectors between grid-points) of the system are 
\begin{equation}
\begin{array}{c}
\textbf{e}_{x} = \frac{1}{RB_{\theta}}\hat{\textbf{e}}_{\psi} + IR\hat{\textbf{e}}_{\zeta}\\
\textbf{e}_{y} = h_{\theta}\hat{\textbf{e}}_{\theta} + \nu R \hat{\textbf{e}}_{\zeta}\\
\textbf{e}_{z} = R\hat{\textbf{e}_{\zeta}}.
\end{array}
\end{equation}
where $R$ is the major radius, $B_{\theta}$ is the poloidal magnetic field strength, $h_{\theta}$ is the poloidal arc length, $\nu$ is the magnetic field line pitch and $I$ is the integrated magnetic shear which will be described subsequently. The $x,z$ plane, in which the perpendicular dynamics predominantly occur (though note that $z$ is not the binormal direction), contains $132\times 128$ grid points whilst the $y$ coordinate, which is entirely parallel to the field line, is discretized with $128$ grid points. In the parallel direction this has a spatial resolution of $13cm$. This was reduced to $1.6cm$ without any change in the observable physics of the filament so the lower resolution grid was chosen due to its computational benefit. Variations in the parallel structure of filaments with drift-waves stabilized occur on the length scale of the equilibrium magnetic field, which is typically on the $m$ scale, so the parallel direction is well resolved. In the $x,z$ plane the grid resolution is typically on the $mm$ length scale, though this is complicated by the squeezing of the grid around the X-points (see figure \ref{grid_spacing}). Interchange behaviour develops on the length scale of the filament cross-section, which is typically on the $cm$ scale which is well resolved by the grid. Resistive drift-waves are most unstable when $k_{\perp}\rho_{s} = 1$, which requires resolution of length scales $L \sim 2\pi \rho_{s}$. This is satisfied for temperatures upwards of $\sim 5eV$. For simulations at temperatures below this threshold the most unstable drift-wave cannot be driven. For the present studies, resistive drift-waves are stabilized, so this is not foreseen as a problem. As a test the blob size was reduced and drift-waves were observed at roughly the threshold predicted by Angus \cite{AngusPRL2012}. Furthermore grid convergence studies were carried out in both the parallel and perpendicular direction and did not reveal any loss of sub-grid scale physics at the grid resolutions presented here, so the grid resolutions are considered adequate for the simulations presented herein.
\\ \\BOUT++ \cite{BOUT++} calculates the metric tensor of the field aligned system (14) internally given a set of equilibrium parameters, shown in figure \ref{Metric_quantities}, which consist of: the magnetic field, $B$, and its poloidal and toroidal components; the integrated magnetic shear $I = \int\nu ' dy$ where $\nu ' = \partial \nu / \partial \psi$ is the local magnetic shear; the major radius, $R$ and vertical distance $Z$ of the field line (in the cylindrical coordinate system). 
These equilibrium parameters have been extracted from an EFIT \cite{EFIT} equilibrium reconstruction of MAST shot 14220 which is an Ohmic double-null L-mode discharge. With this information derivatives in the field aligned system are calculated self-consistently and internally within BOUT++. A corollary of using the field aligned coordinate system is that the extent of the domain in real space varies along the length of the field line. Figure \ref{grid_spacing} shows the size of a grid spacing in $\psi$ and $z$ in real space, denoted as $d\psi$ and $dz$, along the length of the field line.
Figure \ref{grid_spacing} shows that a domain initially stretched in the $z$ direction at the midplane ($y=0m$) becomes stretched in the $\psi$ direction by the time the target plate is reached. The motivation for choosing $d\psi$ and not $dx$ directly is that $d\psi$ is the component of $dx$ perpendicular to the field. Importantly this is not a rotational transform but rather a stretching/contraction transformation in both the $x$ and $z$ directions. This is an unavoidable consequence of the choice of coordinate system used for this investigation. This means that it is impossible for a filament to be seeded homogeneously along the field line in both real space and the field aligned system. For the purposes of this paper filaments are seeded homogeneously in the field aligned system.
\\ \\The polarization vector, $\xi$ defined in (5) and presented in figure \ref{curvature}, is an important geometrical feature of the system since it determines the strength of drive for the interchange mechanism.
It defines the direction of polarization across the filament due the curvature drift. The propagation direction of the filament due to the interchange mechanism is then the direction normal to $\xi$ and the magnetic field. $\xi^{z}$ is the dominant component at most points along the field line, apart from in the vicinity of the X-point (around $y\sim \pm 5m$). In this region the curvature changes sign and becomes comparable in magnitude with $\xi^{x}$ (ignoring $\xi^{y}$ since the dynamics occur in the perpendicular plane only). The components of $\xi$ shown in figure \ref{curvature} are derived from the same EFIT equilibrium.
\\ \\Boundary conditions for the simulations presented in this paper are Neumann in the $x$ direction and periodic in the $z$ direction with a period $P=30$. The simulation is formally an $n=30$ distribution of filaments. This is representative of the experimental distribution of filaments, which are observed in MAST with an average toroidal mode number $n\sim 30-50$\cite{DudsonPPCF2008}. The periodicity of the filaments was varied, with filament size remaining constant and no change in the physics of the filaments was observed which allows the simulations to be considered as isolated filaments in that there is no interaction with any neighbouring filament.  The boundary conditions in $y$ are Neumann for all variables apart from the parallel current. On the divertor plate boundaries the parallel current is matched to the sheath current \cite{NedospasovNF85} such that
\begin{equation}
J_{||} = enc_{s}\left(1 - \exp\left(-\phi\right)\right)
\end{equation}
on the upper plate and :
\begin{equation}
J_{||} = enc_{s}\left(1 - \exp\left(\phi\right)\right)
\end{equation}
on the lower plate where $\phi$ and $n$ are the boundary values. Finally the simulations are initialised with a Gaussian density perturbation on the $x,z$ plane which is homogeneous along $y$. The time-integration is performed by an implicit Jacobian-free Newton-Krylov solver which has adaptive time-stepping ensuring that all relevant time-scales are included in the solution. Simulations dominated by interchange dynamics took $\sim 12$ hours on 8 cores, whilst simulations dominated by Boltzmann dynamics required much higher time resolution and took $\sim 36 - 168$ hours on 8 cores. Spatial derivatives are solved with $4th$ order central differencing in $x$ and $y$ whilst FFTs are used in the periodic $z$ direction.
\\ \\All filament simulations within this paper have identical initial conditions. In all cases $\delta n/n \approx 1.6$, $\delta_{\perp}\approx 3cm$ (filament radius) at the mid plane and the filament is initially homogeneous along the field line in the $(x,y,z)$ system connecting from divertor plate to divertor plate with a flat background density and no other background variables. The filament potential is not seeded, but develops a polarization very quickly as the simulation progresses. From Angus \emph{et.al} \cite{AngusPRL2012,AngusPoP2012} the condition for drift-waves to impact blob dynamics is $0.15\sqrt{R_{c}/\delta_{\perp}}\geq 1$ where $R_{c}$ is the radius of curvature. Given these initial conditions and the magnetic parameters detailed in figures \ref{Metric_quantities} and \ref{curvature},  $0.15\sqrt{R_{c}/\delta_{\perp}} = 0.87 < 1 $ indicating that the dynamics should remain relatively unaffected by drift-waves. This is helpful given the desire to study the transition from interchange to Boltzmann dynamics without the added complication of unstable resistive drift-waves.

\section{Effects of magnetic geometry on filament structure}
 
\begin{figure}[ht!]
\centering
\includegraphics[width=0.6\textwidth]{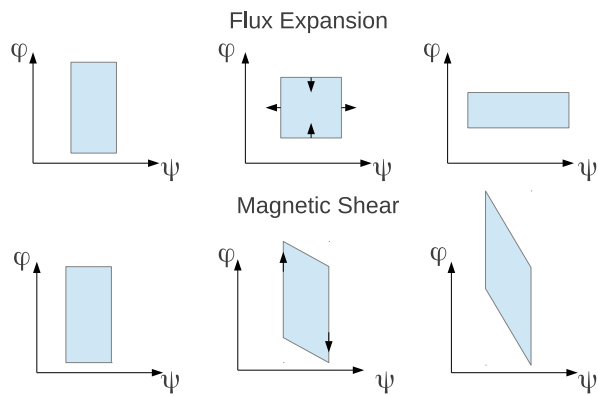}
\caption{Schematic describing the effects of flux expansion and magnetic shear on an initially rectangular cross-section. }
\label{f_expand_shear_schematic}
\end{figure}

\begin{figure}[ht!]
  \begin{center}
    \subfigure[Curvature variation]{
      \includegraphics[width=0.3\textwidth]{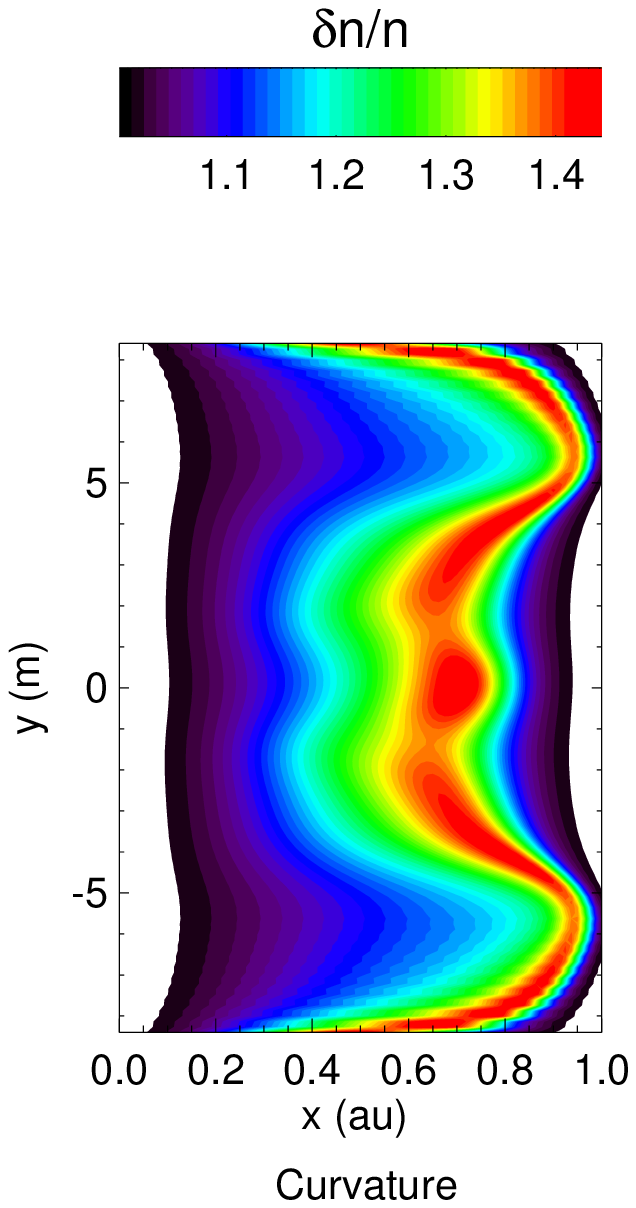}
      \label{par_struct:curv}
      }
    \subfigure[Flux expansion]{
      \includegraphics[width=0.3\textwidth]{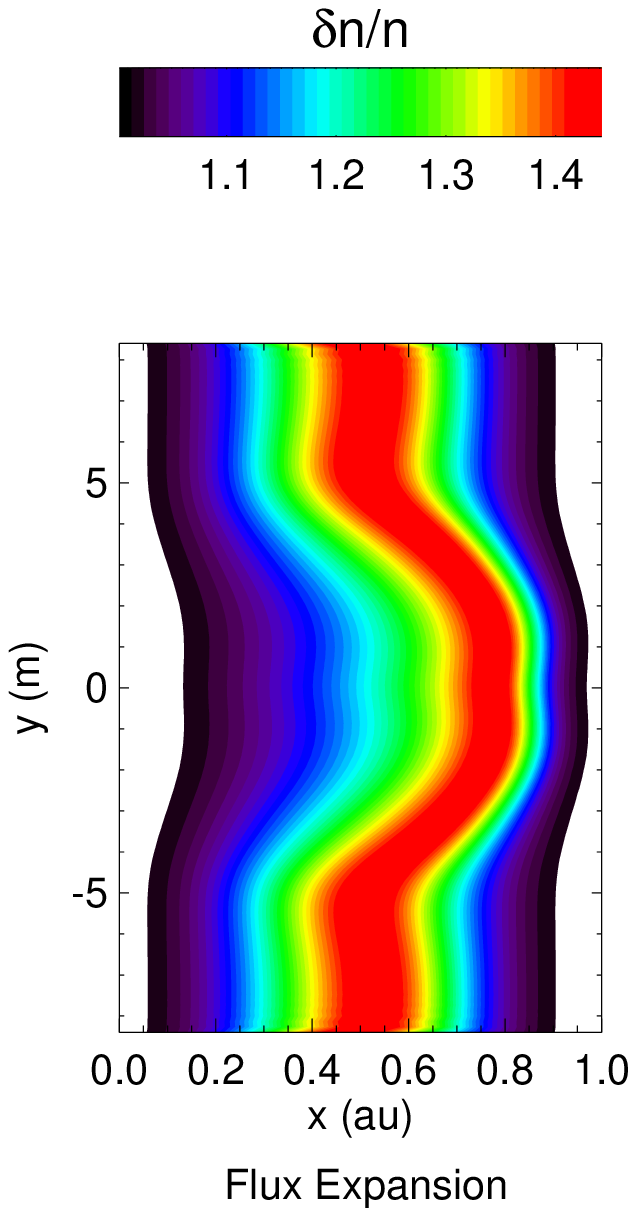}
      \label{par_struct:FX}
      }
    \subfigure[Shear]{
      \includegraphics[width=0.3\textwidth]{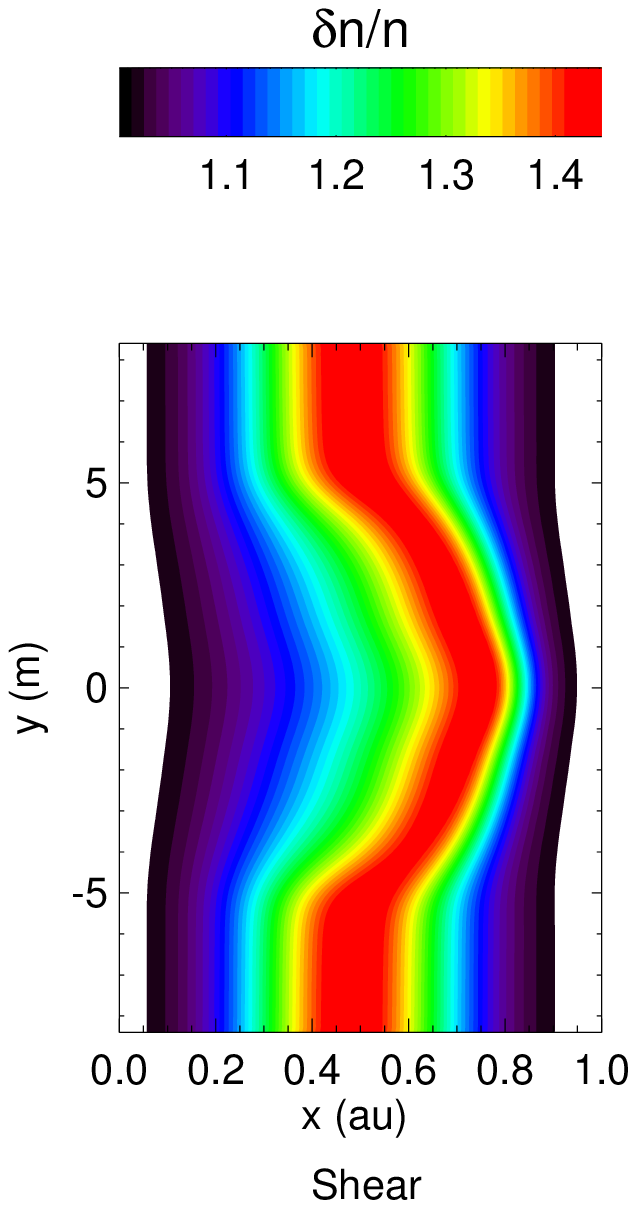}
      \label{par_struct:shear}
      }
    \end{center}
  \caption{Parallel profile of a filament including in each case only one factor driving the parallel structure. Data was sampled at $0.5ms$ in each case. All simulations were otherwise identical with $T_{e} = 1eV$ and $n_{0} = 10^{19}m^{-3}$.}
\label{par_struct}
\end{figure}

\begin{figure}[htp]
\begin{center}
  \subfigure[With sheath conditions]{
    \includegraphics[width=0.4\textwidth]{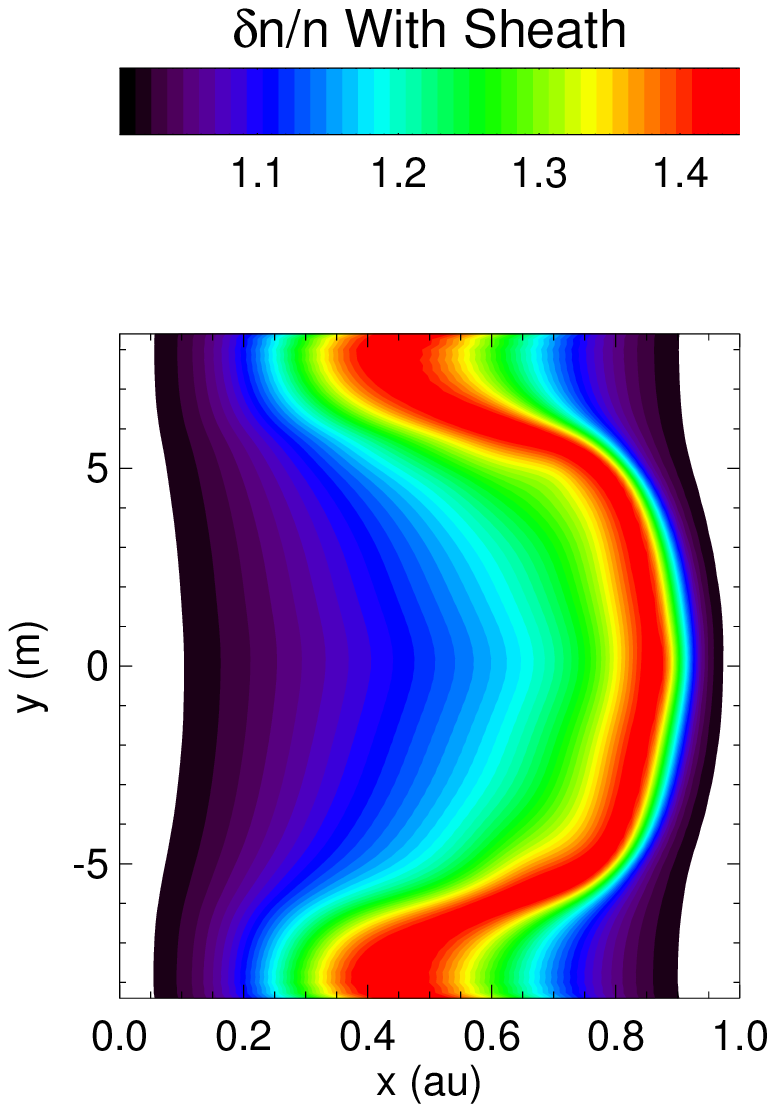}
    \label{div_iso:sheath}
    }
  \subfigure[Without sheath conditions]{
    \includegraphics[width=0.4\textwidth]{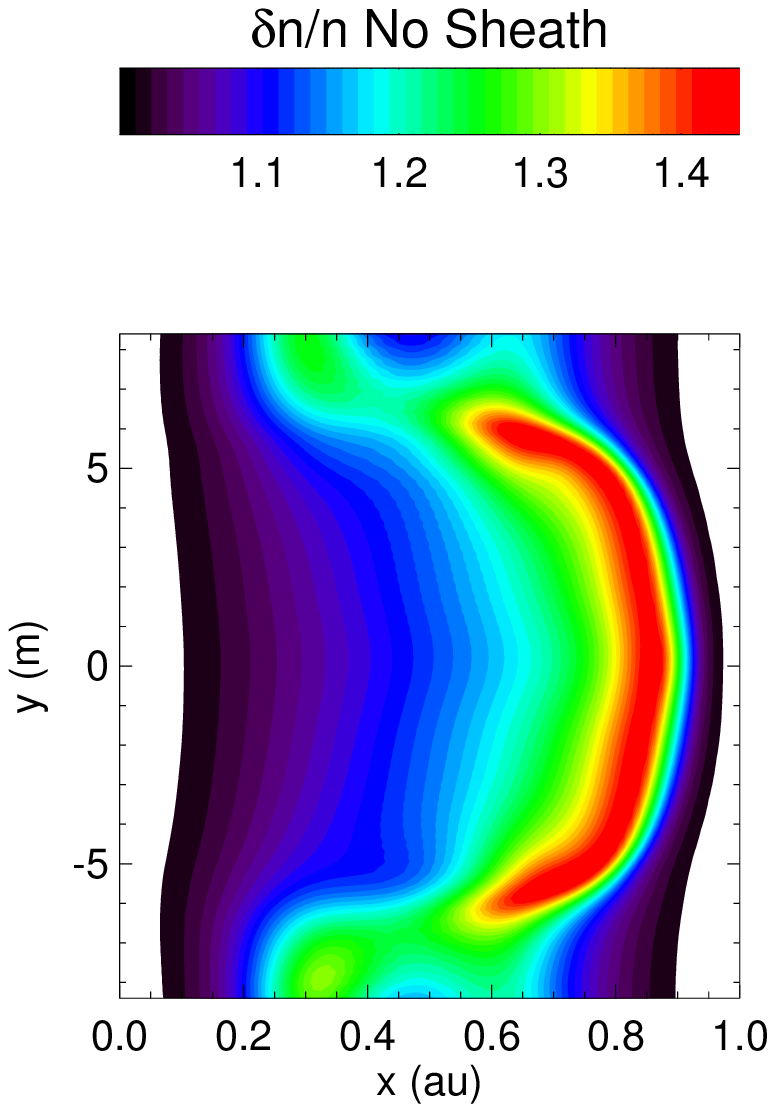}
    \label{div_iso:no_sheath}
    }
  \end{center}
\caption{Parallel profiles of filament simulations with and without sheath boundary conditions included. In the case without sheath boundary conditions Neumann conditions were used on $J_{||}$ instead . The effect of the sheath is clearly localised to the divertor region. Data was sampled at $0.2ms$ and simulations are otherwise identical.}
\label{div_iso}
\end{figure}

\begin{figure}[ht!]
\includegraphics[width=0.35\textwidth]{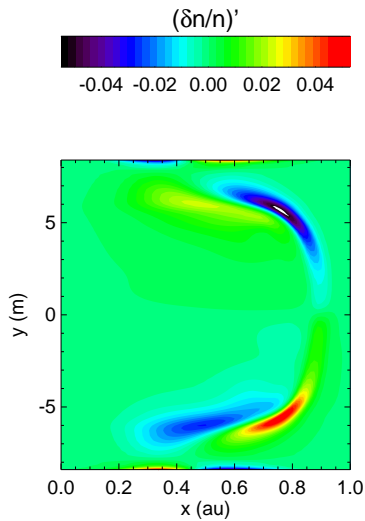}
\centering
\caption{Parallel gradients in the flux tube as a result of the divertor regions independance from the misplane. The gradients are strongest around the X-point, but persist into the midplane with a smaller magnitude. Data is taken from the same simulation and at the same time as in figure \ref{div_iso:sheath}.}
\label{par_grad}
\end{figure}

In the conventional 2D theory of blobs where interchange motion drives the propagation of the filament \cite{KrasheninnikovPhysLet2001,YuPoP2006,GarciaPoP2006,MyraPoP2004, YuPoP2003,D'IppolitoPoP2004} , the cross-field motion is independent of the 3D structure of the filament. However modelling including 3D effects\cite{AngusPRL2012,AngusPoP2012} has shown that accounting for the full 3D structure of the filament can become important. Since only the \emph{local} drive for the interchange mechanism determines the local propagation velocity of the filament, any variation in the drive along the field line will lead to a variation in the propagation velocity along its length, giving rise to a 3D structure. In the model presented here there are three factors which affect the drive for the interchange motion:
\begin{enumerate}
\item The variation of the polarization vector magnitude, $\left|\xi\right|$, along the field line,
\item The change in size of the flux tube due to flux expansion,
\item The twisting of the system due to magnetic shear.
\end{enumerate}
The strength of the polarization vector determines the strength of the polarized electric field which leads to the filaments advection. In regions where the magnetic curvature is strong, $\xi$ is large and the filament advects outwards quickly. By contrast when the curvature is weak, $\xi$ is small and the filament advects slowly. This variation in the advection speed is what gives rise to the 3D structure. 
\\Flux expansion near the X-point squeezes the flux tube \cite{FarinaNF1993} and leads to highly anisotropic cross sections. This is borne out in these simulations by the change in domain size along the flux tube indicated by figure \ref{grid_spacing}. This squeezing of the flux tube can stabilize the linear interchange instability when the dimensions of the flux tube become comparable to the ion gyro-radius, $\rho_{s}$\cite{FarinaNF1993,RyutovConPlasmaPhys2008}. In the filament simulations conducted here the squeezing effect was not severe enough to stabilize the interchange instability, however  the strength of the non-linear interchange mechanism that drives 2D filament motion is drastically reduced in the highly squeezed region near the X-point \cite{RyutovConPlasmaPhys2008,RyutovConPlasmaPhys2006}. Physically this reduction in drive is attributed to a greater spreading of polarized charge in the X-point region, which leads to a drop in the polarization electric field and consequently reduced advection rates. The filament is therefore strongly driven near the midplane, where the $z$ dimension is un-squeezed, and weakly driven near the X-point where the $z$ dimension is heavily squeezed. The magnetic shear also spreads the charge over a larger region by shearing the filament. In this case the width in $\psi$ is unchanged however the $x$ coordinate develops a component in the toroidal direction. This reduction in drive by squeezing the filament leads to the formation of a parallel structure to the filament. The effects of flux expansion and magnetic shear are illustrated (in an exaggerated manner) in figure \ref{f_expand_shear_schematic}. 
The magnetic shear also affects the curvature. The driving component of magnetic curvature is in the $\nabla\psi$ direction ($\xi^{\perp}$ in figure \ref{curvature:psi}). As the magnetic shear gets stronger this develops a strong component in $\xi^{x}$ which then varies the direction of propagation of the filament in the $(x,z)$ plain to ensure that the dominant advection direction is in $\nabla\psi$. This gives rise to a twisting of the filament as it passes the X-point in the $(x,y,z)$ system. 
\\Figure \ref{par_struct} shows three simulations which exhibit the appearance of a parallel structure to the filament.
In each case the filament was driven purely by interchange motion (to decouple the complex effects of 3D dynamics from the effects of magnetic geometry). In the first simulation only the variation in drive along the flux tube due to magnetic curvature was included. In the second only the effects of flux expansion were included and in the final panel only the effects of the integrated magnetic shear were included. The magnetic curvature variation leads to a ballooning of the filament in the region of strongest curvature (i.e. largest $\xi^{\perp}$). This effect is greatest near the X-points, where the magnetic field becomes very curved. Both the flux expansion and the magnetic shear cases show an opposite trend with the filament ballooning at the midplane. This is the region where the integrated shear is $\sim$ 0 (since the integration was centred on the mid plane) and where the filament is strongly stretched in the $z$ direction. These two effects serve make the dynamics of the filament at the midplane independent from the divertor region. In figure \ref{div_iso} two simulations are presented showing the parallel structure that develops in a filament including all of the effects highlighted above. 
Figure \ref{div_iso:sheath} shows a full filament simulation with sheath boundary conditions applied to $J_{||}$ at the boundaries in $y$ (ie approximately at divertor target plates). The midplane portion of the filament (the part between X-points) is noticably ballooned with respect to the divertor portion. This signifies that the effects of flux expansion and magnetic shear supersede the curvature drive in the formation of parallel density gradients due to varying interchange drive along the filament. The simulations are performed with parameters taken from the far SOL in MAST. It is plausible that closer to the seperatrix, where curvature will be a lot stronger around the X-points, the ballooning feature near the X-points may become prevalent. An investigation of this requires a fully 3D representation of the magnetic geometry which is beyond the scope of this paper. 
\\ \\It is known that in 3D filaments the effects of the sheath on the cross-field dynamics decrease with distance from the sheath \cite{JovanovicPoP2008}. This has been tested here by rerunning the simulation in figure \ref{div_iso:sheath} but replacing the sheath boundary conditions on $J_{||}$ with standard neumann conditions to preserve $\nabla\cdot J = 0$. In the midplane region the simulations remain qualitatively and quantitatively indistinguishable. Below the X-points in the divertor region the sheath boundary conditions limit the advection velocity of the filament and a noiticable change in the cross-field structure of the filament occurs between the two simulations, however this change in structure is difficult to quantify due to the effects of flux expansion and magnetic shear. It is clear that the X-points screen filament in the midplane from the sheath and make the filament in the midplane region independent of conditions below the X-point. This independance from the sheath has been noted in the two-region model of Myra \emph{et.al} \cite{MyraTwo-Region1} where the effects of flux expansion are the dominant cause. This supports the argument \cite{RussellPRL2004} that filaments in the SOL of a tokamak depend on the boundary conditions at the X-point(s) rather than at the divertor plates. It has been shown \cite{RyutovConPlasmaPhys2008} that the filament can re-couple to the sheath as it propagates if the decoupling effects of flux expansion and magnetic shear weaken during the filaments outward motion. This does not occur in the simulation presented here since no cross-field variations in magnetic parameters have been included. The decoupling from the sheath means that the build up of charge due to drift motion can only be dissipated inertially. This is the familiar resistive ballooning model of filament motion \cite{MyraTwo-Region1} which gives faster propagation velocities than the sheath-limited model and is the reason that the midplane region advects more quickly than the divertor region. Within the two-region model the parameters $\Lambda = L_{||}\left(\tau_{e}\Omega_{e}\rho_{s}\right)^{-1}$ and $\Theta = \left(\delta_{\perp}/a*\right)^{5/2}$ where $a* = (L_{||}^{2}\xi/(2\rho_{s}))^{1/5}$ \cite{MyraTwo-Region1} determine the dynamics of the filament. Taking midplane parameters from the simulations conducted in this section gives $\Lambda \sim 10^{2}$ and $\Theta \sim 10^{-2}$. The condition for a blob to be in the resistive ballooning regime is that $\Lambda>\Theta$\cite{MyraTwo-Region1}. It is therefore highly likely that filaments in MAST are resistive ballooning rather than sheath limited. In the two region model the resistive ballooning blob is decoupled from the sheath boundary, in agreement with observations from the simulations presented here (figure \ref{div_iso}). Despite this decoupling, unless otherwise stated, all simulations performed in this paper employ the sheath boundary conditions since these represent the closest approximation to reality available within the model. The fact that the midplane region of the filament is limited by inertia rather than the sheath dissipation means that the filament can achieve greater velocities than in the divertor region. As such particles from the filament tracked along a field line until intersection with the divertor target may do so at a greater radius than expected if one only considers the sheath limited model. Furthermore this suggests that a change in boundary conditions near the divertor target plate, going to a detached divertor regime for example, will not significantly affect filament motion around the midplane. This would be a good topic for an experimental comparison.  It should be noted that the effects of parallel streaming are not observed in these simulations due to the neglect of parallel ion dynamics in the governing equations.
\\ \\The effects that contribute to the dynamic independance of the midplane filament are strongest in the vicinity of the X-point. Figure \ref{par_grad} shows the parallel density gradients, $\frac{\partial n}{\partial y}$ that arise around the X-point. 
These density gradients arise as a consequence of the magnetic geometry and are therefore inevitable in a realistic tokamak scenario. As has been indicated earlier, parallel density gradients can become the dominant term in parallel Ohm's Law, (4), which can drive the filament towards a Boltzmann response. In the next section the transition to the Boltzmann regime is investigated.

\section{3D effects on cross-field motion}

\begin{figure}[ht!]
\begin{center}
  \subfigure[]{
    \includegraphics[width=0.45\textwidth]{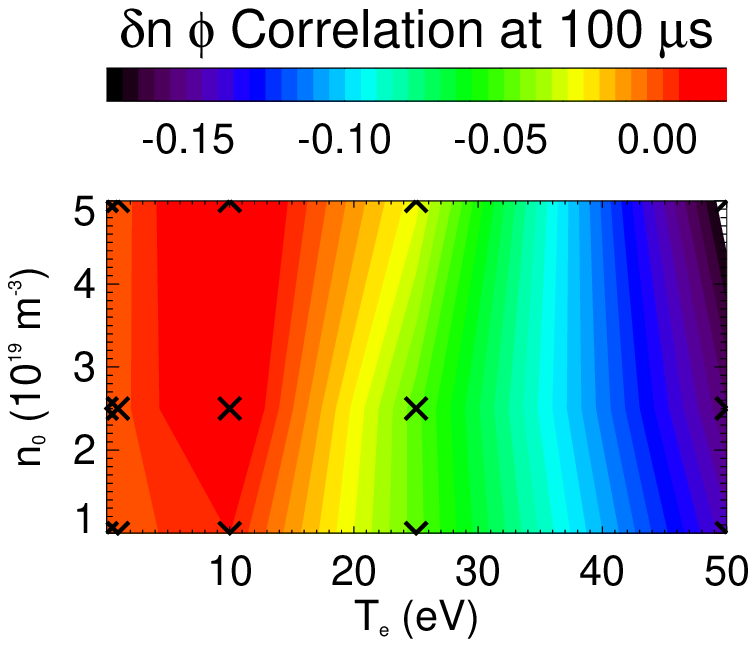}
    \label{corr_contour:abs_time}
    }
  \subfigure[]{
    \includegraphics[width=0.45\textwidth]{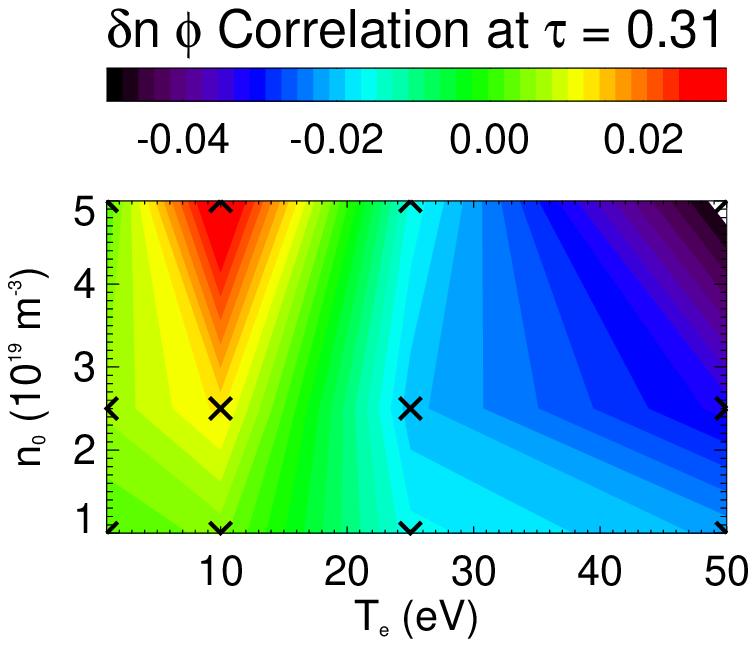}
    \label{corr_contour:cs}
    }
  \end{center}
\caption{Interpolated contour plot of the correlation parameter $C$ in $n_{0}$ $T_{e}$ space sampled at $t=100\mu s$ (a) and $\tau = \Delta t c_{s}/L_{||} = 0.31$ (b). $\tau = 0.31$ is chosen since it allows the filament motion to develop sufficiently, whilst still allowing neglect of parallel ion dynamics. A transition occurs with increasing $T_{e}$, as predicted, however the predicted transition with decreasing $n_{0}$ is not observed. This may require simulations of densities lower than $1\times 10^{19}m^{-3}$. Crosses indicate the simulated points that comprise the contour. All filaments are identical in all other aspects.}
\label{corr_contour}
\end{figure}

Normalizing equation (13) in the gyro-Bohm convention, i.e. normalizing time-scales by $1/\Omega_{i}$, length scales by $\rho_{s}$ gives and densities by $n_{0}$, the background density, gives
\begin{equation}
\frac{\partial \nabla_{\perp}^{2}\phi}{\partial t} =  2\hat{\xi}\cdot\frac{\nabla n}{n} - \left[\textbf{b}\times\nabla\phi\cdot\nabla\right] \nabla_{\perp}^{2}\phi + \frac{\alpha}{n}\left(\nabla_{||}\left(\frac{\nabla_{||}n}{n}\right) - \nabla_{||}^{2}\phi\right)
\end{equation}
where the dimensionless parameter $\alpha$ is  
\begin{equation}
\alpha = \frac{T_{e}\sigma_{||}}{\rho_{s}n_{0}c_{s}e^{2}} = \frac{\sigma_{||}B}{en_{0}}
\end{equation}
and $\hat{\xi} = \rho_{s}\xi \sim \rho_{s}/R_{c}$ is the dimensionless curvature. The physics of filament propagation is an advection of density by an $\textbf{E}\times\textbf{B}$ velocity field. The form that the advection velocity takes is dependant on the collisionality. In the initial stages of growth, after the seeding of the filament, the non-linear self-advection term (2nd RHS term) is small and since the filaments are initially homogeneous along the field line the parallel gradient terms are also small. The only term left is the interchange term which correlates the perpendicular derivative of the potential with the density perturbation. If the initial density perturbation has any symmetry then this correlation ensures that the potential structure has symmetry of exactly opposite parity; this is the familiar formation of dipolar potential lobes around the density perturbation. As has already been discussed, the interchange mechanism then leads to the formation of parallel gradients in density and potential due to the magnetic geometry. The sustained parallel density gradients drive parallel currents by parallel Ohm's law, given here in normalized form
\begin{equation}  
\eta_{||}J_{||} = \frac{T_{e}}{n_{0}c_{s}\rho_{s}e^{2}}\left(\frac{\nabla_{||}n}{n} - \nabla_{||}\phi\right)
\end{equation}
where $\eta_{||}$ is the parallel resistivity. At high collisionality $\eta_{||}$ is large which allows a significant phase offset between density and potential perturbations and can drive unstable drift waves. In the simulations investigated here the filament radius, which is $\delta_{\perp} \sim 2cm$ at the midplane, is chosen such that the interchange instability outgrows unstable drift-waves. In these conditions Angus has shown that drift-waves do not affect filament dynamics \cite{AngusPRL2012,AngusPoP2012}. It should be noted that the calculations of Angus \emph{et.al} are made in a shear-less slab geometry. This neglects the stabilizing effect of magnetic shear and any other effects associated with magnetic geometry. Since the simulation domain here is local in the sense that there is no cross-field variation in any parameters, the results of Angus can be applied, however the extension of such calculations to a global tokamak geometry would be a good topic for future research. When collisionality is high $\alpha$ is small and the interchange terms continue to dominate the vorticity equation. By contrast when collisionality is low $\eta_{||}$ is small and $\nabla_{||}\ln\left( n \right) \sim \nabla_{||}\phi$. The potential now begins to align with the density as the plasma tends towards a Boltzmann response. The potential forms a monopolar perturbation which acts to spin the filament about its center. This is exactly the process of Boltzmann spinning described in \cite{AngusPoP2012} however in this work the parallel density gradients which drive the process arise as a natural consequence of the geometry and are not imposed as a starting condition. Since charge conduction along the field line is much faster when the Boltzmann response is driven, charge polarization cannot build up and the spinning motion due to the Boltzmann response is comparable to the ejective motion which leads to a net spinning of the filament, rather than an expulsive motion radially outwards. 
\\ \\Returning to equation (18), when the non-linear advection term is small the dynamics of the system are governed by $\hat{\xi}$ and $\alpha$. In practice magnetic parameters are approximately fixed for a given magnetic confinement device. Furthermore filament parameters are dependant on the turbulent conditions within the edge and cannot in principle be manually manipulated. The plasma parameters that are practically variable from shot to shot in a tokamak are temperature and density (though such variation is certainly no simple procedure). The dynamics of filaments have been investigated with varying temperature and density here by running otherwise identical simulations over an $n_{0},T_{e}$ parameter space. The filaments have $\delta n/n_{0} = 0.6$, $\delta_{\perp}\approx 2cm$ at the mid plane (here $\delta_{\perp}$ is perpendicular filament radius) and are initially connected from divertor plate to divertor plate. Figure \ref{corr_contour} shows a measurement of the correlation parameter 
\begin{equation}
C = \frac{\langle \phi\delta n \rangle}{\langle\left|\phi\right|\rangle\langle\delta n\rangle}
\end{equation}
across the parameter space, where the angled brackets represent a volume average such that $\langle A \rangle \equiv \int JAd^{3}\textbf{R}$ where $J = h_{\theta}/B_{\theta}$ is the Jacobian of the coordinate system defined by (15). 

\begin{figure}[ht!]
\begin{center}
  \subfigure[]{
    \includegraphics[width=0.3\textwidth]{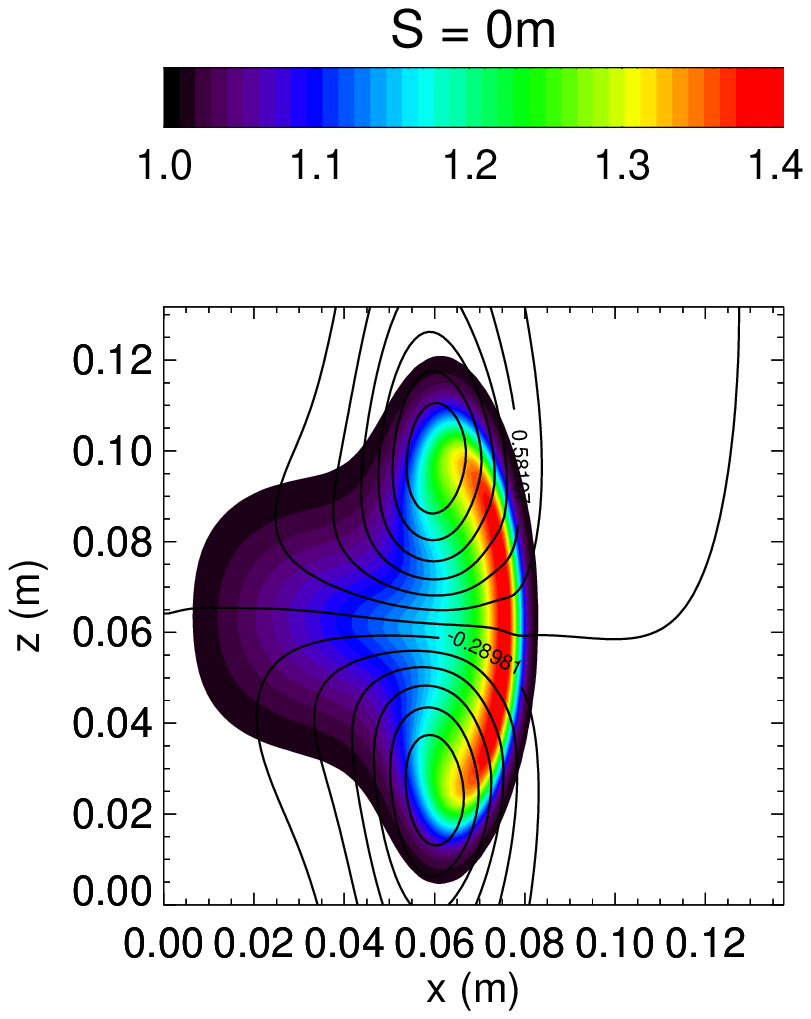}
    \label{int_xsec:mid}
    }
  \subfigure[]{
    \includegraphics[width=0.3\textwidth]{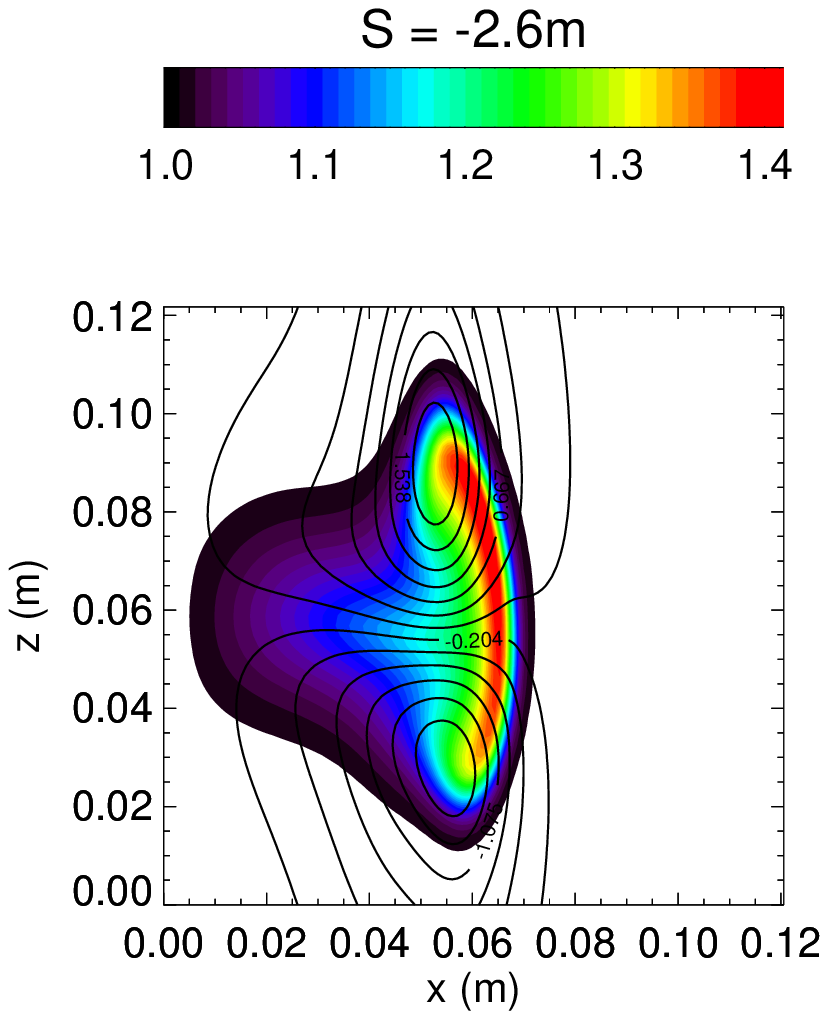}
    \label{int_xsec:lower}
    }
  \subfigure[]{
    \includegraphics[width=0.3\textwidth]{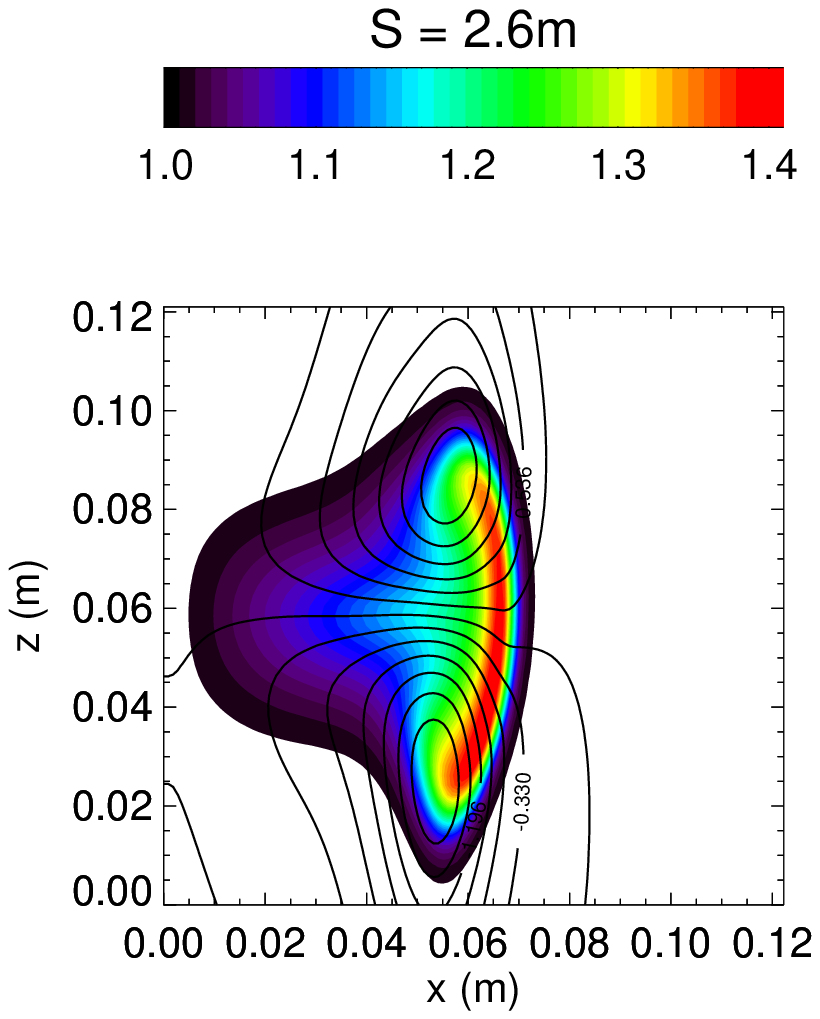}
    \label{int_xsec:upper}
    }
\end{center}
\caption{Cross sections of the filament in the interchange regime at the midplane and $2.6m$ either side of the midplane. Snapshots are taken at $t = 200\mu s$ with $n_{0} = 5\times 10^{19}m^{-3}$ and $T_{e} = 1eV$ placing the filament firmly in the interchange regime. Colour indicates $\delta n/n_{0}$ and contour lines show the electrostatic potential, $\phi$.}
\label{int_xsec}
\end{figure}

\begin{figure}[ht!]
  \begin{center}
    \subfigure[$1\mu s$: Initial interchange behaviour]{
      \includegraphics[width=0.35\textwidth]{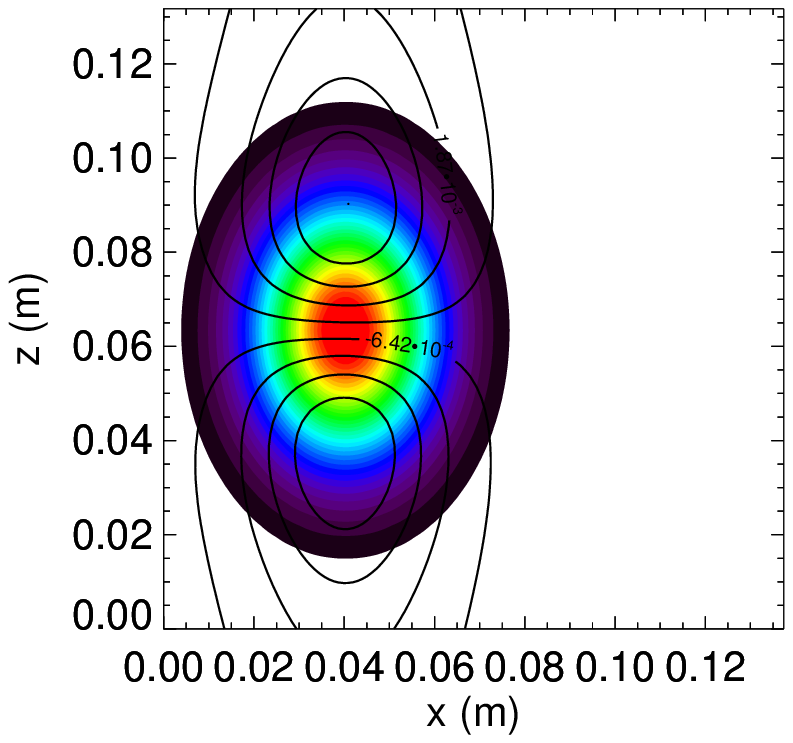}
      \label{Boltz_xsec:1}
      }
    \subfigure[$50\mu s$: Dipole rotation]{
      \includegraphics[width=0.35\textwidth]{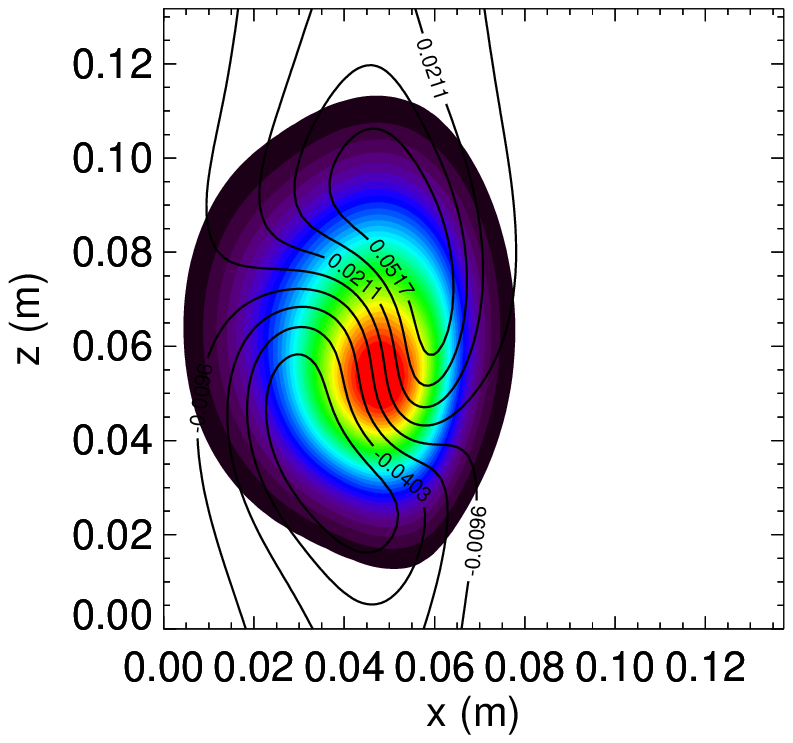}
      \label{Boltz_xsec:2}
      }
    \subfigure[$100\mu s$: Phase matching ]{
      \includegraphics[width=0.35\textwidth]{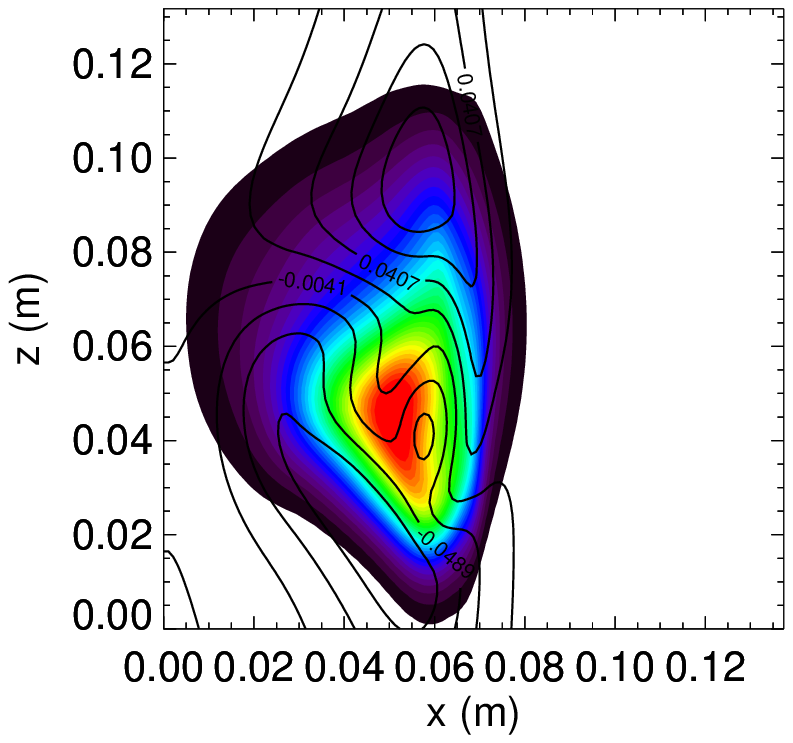}
      \label{Boltz_xsec:3}
      }
    \subfigure[$150\mu s$: Elongation in $z$]{
      \includegraphics[width=0.35\textwidth]{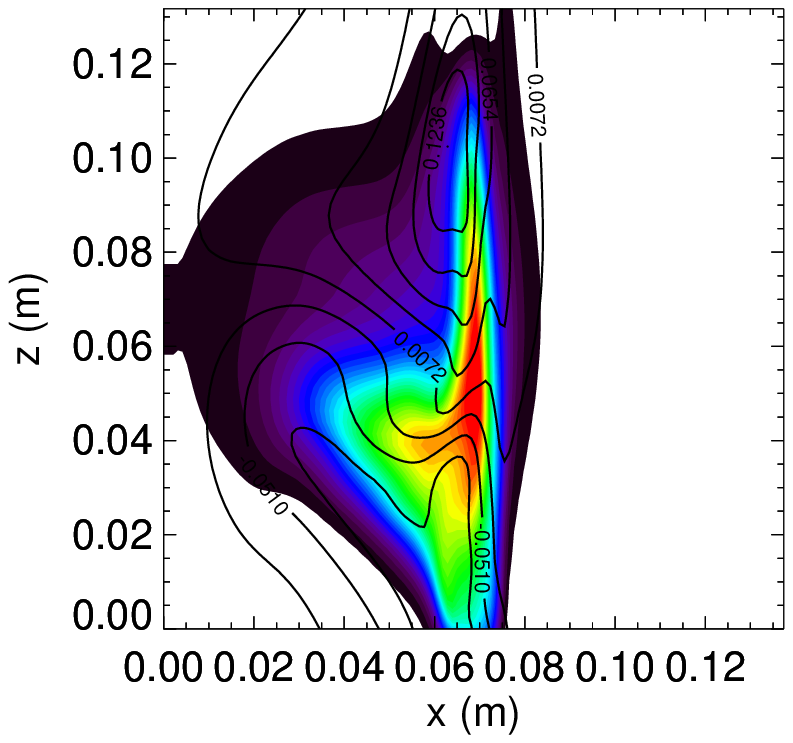}
      \label{Boltz_xsec:4}
      }
    \end{center}
\caption{Cross-field evolution at the midplane of a filament in the Boltzmann regime showing distinctive features of its evolution including a self-organisation due to the Boltzmann response (c) followed by a descent into a turbulence. The simulation was at $T_{e} = 50eV$ and $n_{0} = 2.5\times 10^{19}m^{-3}$.}
\label{Boltz_xsec}
\end{figure}

\begin{figure}[ht!]
\begin{center}
  \subfigure[$1\mu s$]{
    \includegraphics[width=0.2\textwidth]{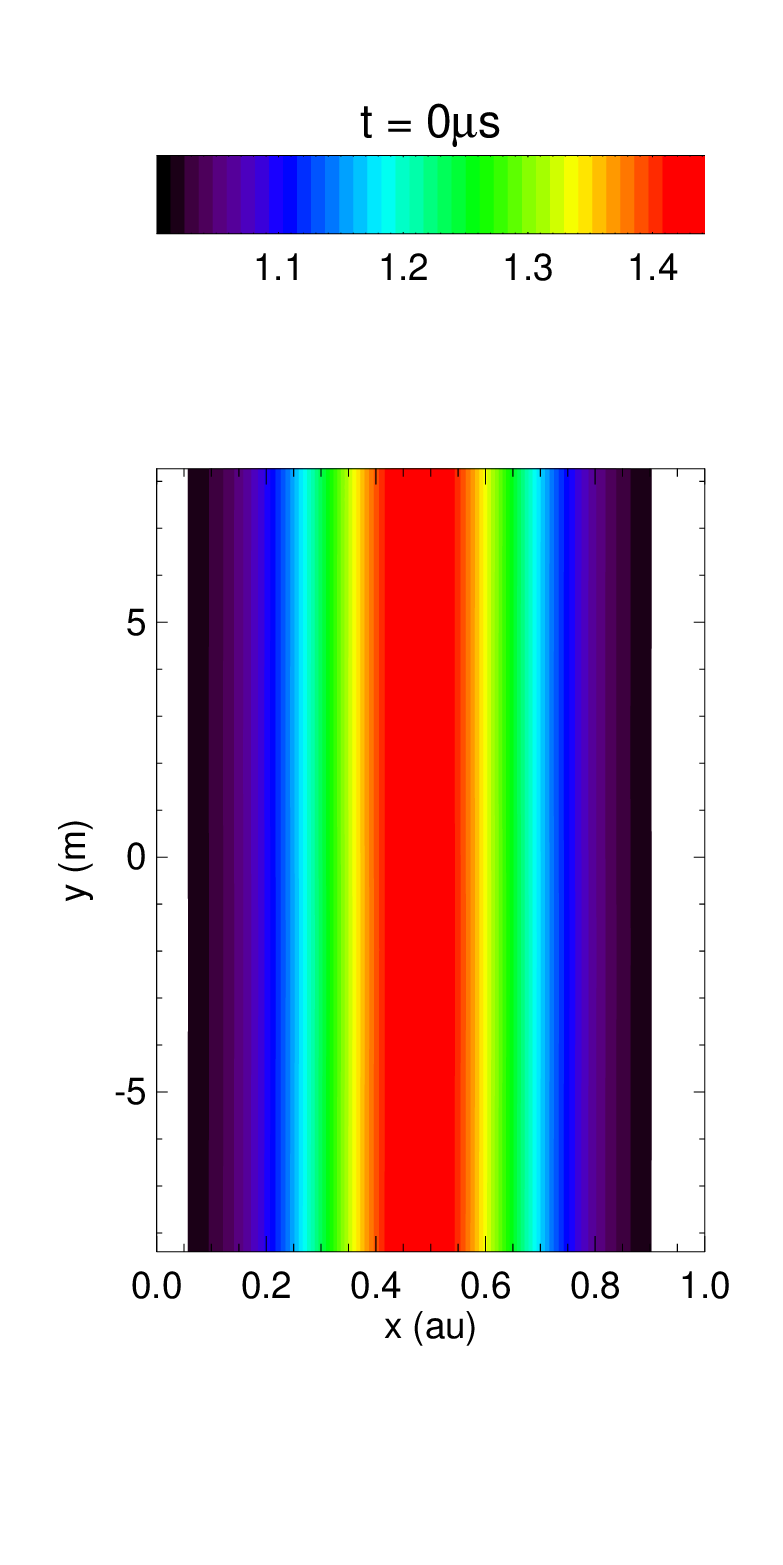}
    \label{Boltz_par:1}
    }
  \subfigure[$50\mu s$]{
    \includegraphics[width=0.2\textwidth]{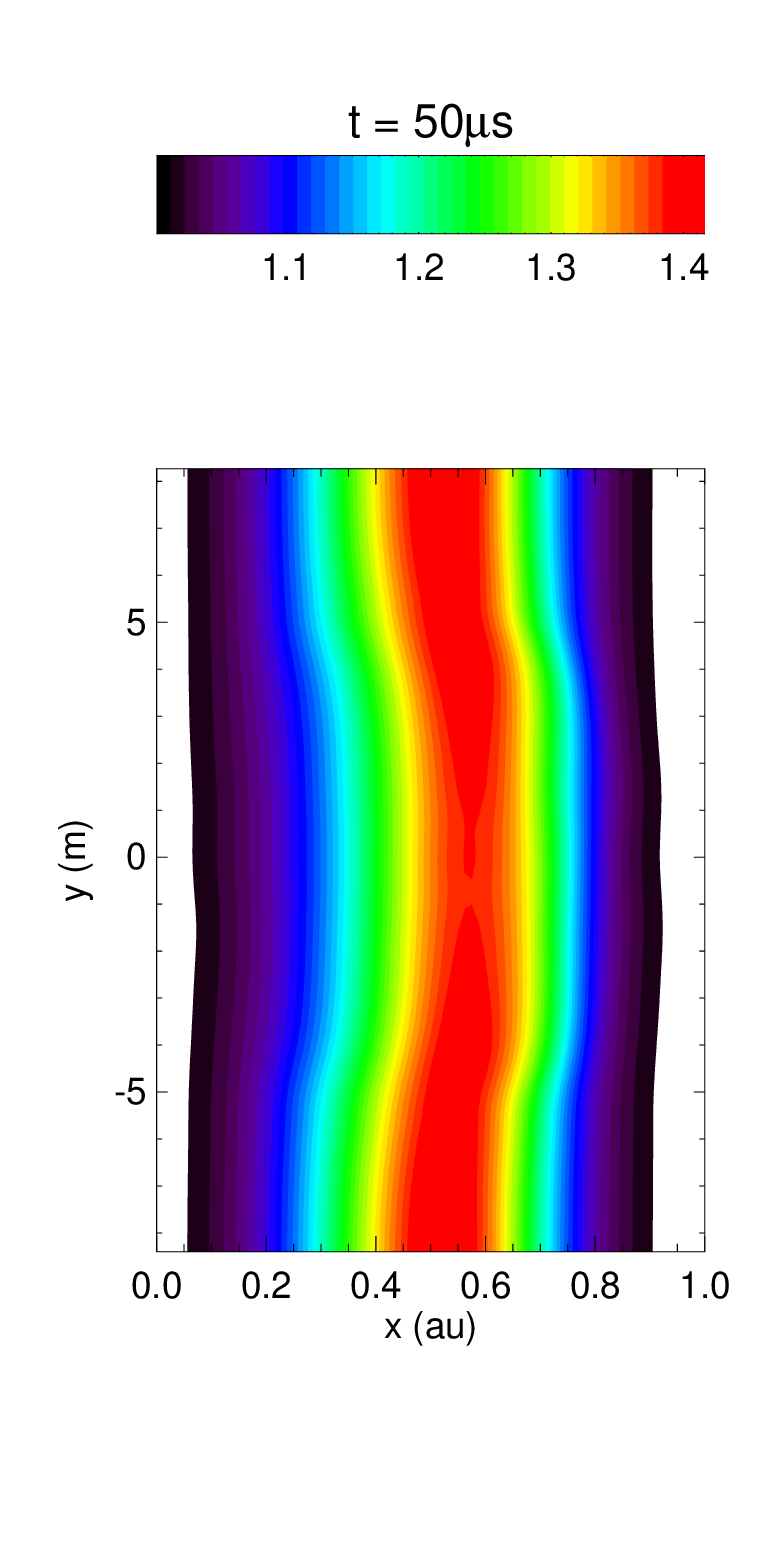}
    \label{Boltz_par:2}
    }
  \subfigure[$100\mu s$]{
    \includegraphics[width=0.2\textwidth]{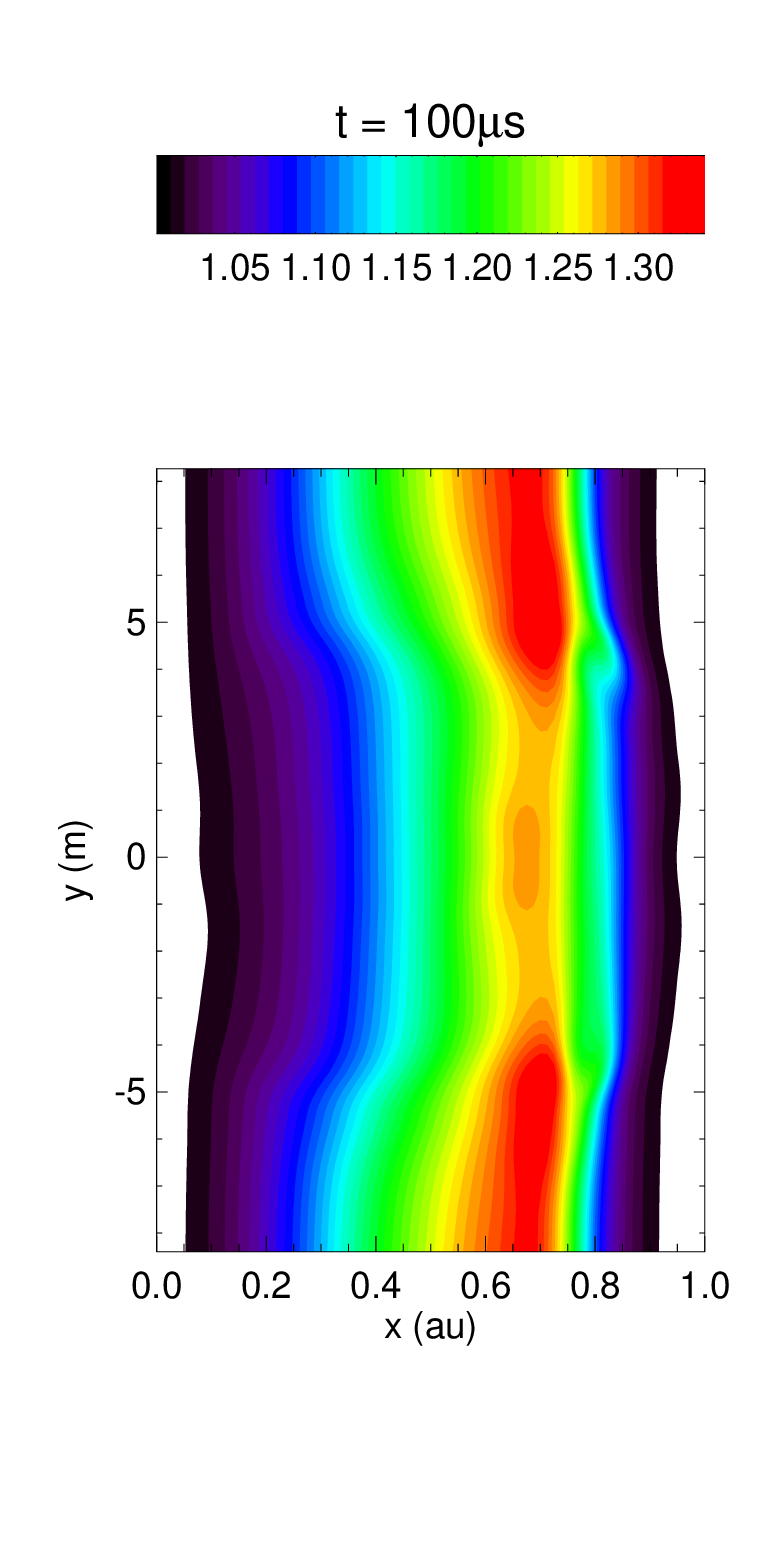}
    \label{Boltz_par:3}
    }
  \subfigure[$150\mu s$]{
    \includegraphics[width=0.2\textwidth]{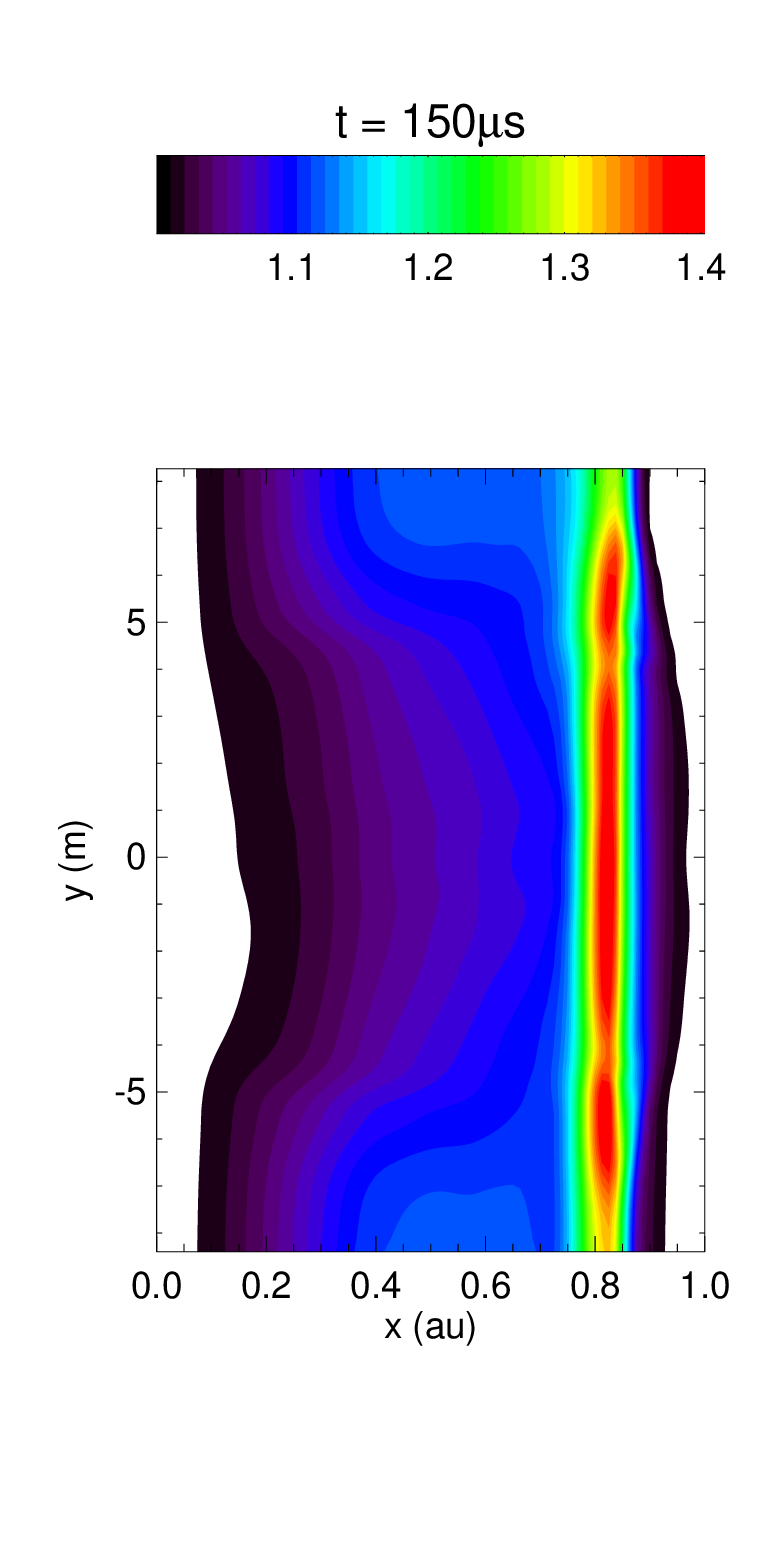}
    \label{Boltz_par:4}
    }
  \end{center}    
\caption{Parallel profile of a filament in the Boltzmann regime. The filament remains connected to both sheaths and does not exhibit as much parallel structure as observed in the interchange regime. Data is from the simulations presented in figure \ref{Boltz_xsec}.}
\label{Boltz_par}
\end{figure}

\begin{figure}
\begin{center}
  \subfigure[]{
    \includegraphics[width=0.3\textwidth]{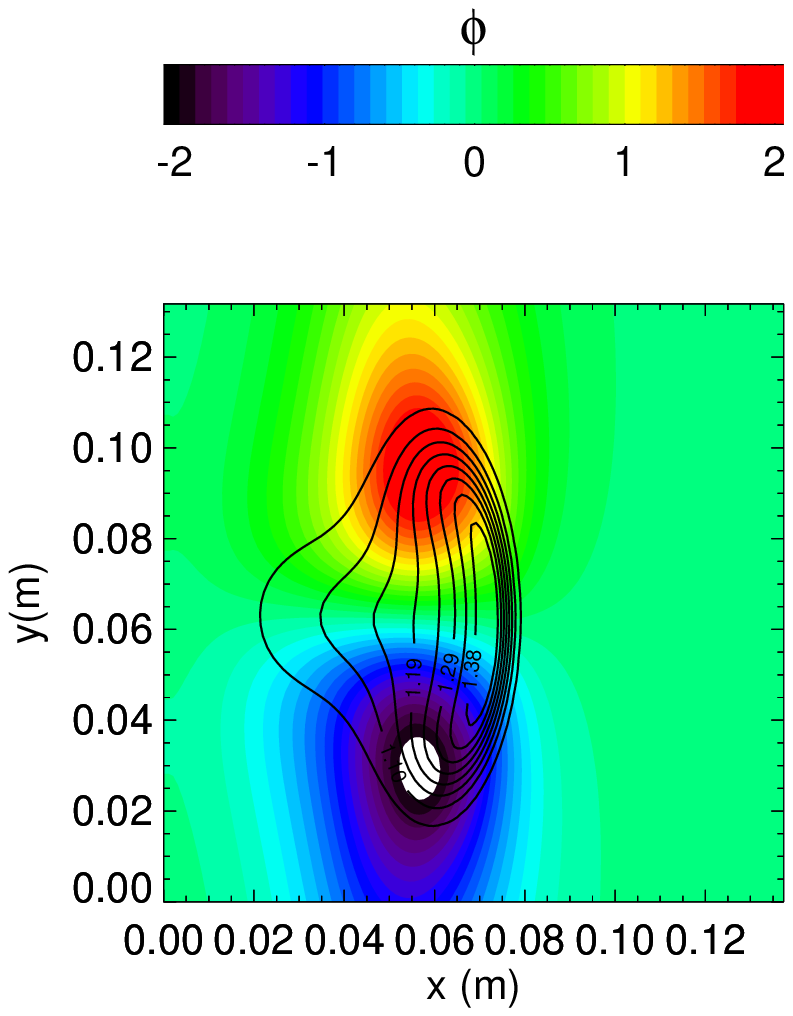}
    \label{morphology:int_phi}
    }
  \subfigure[]{
    \includegraphics[width=0.3\textwidth]{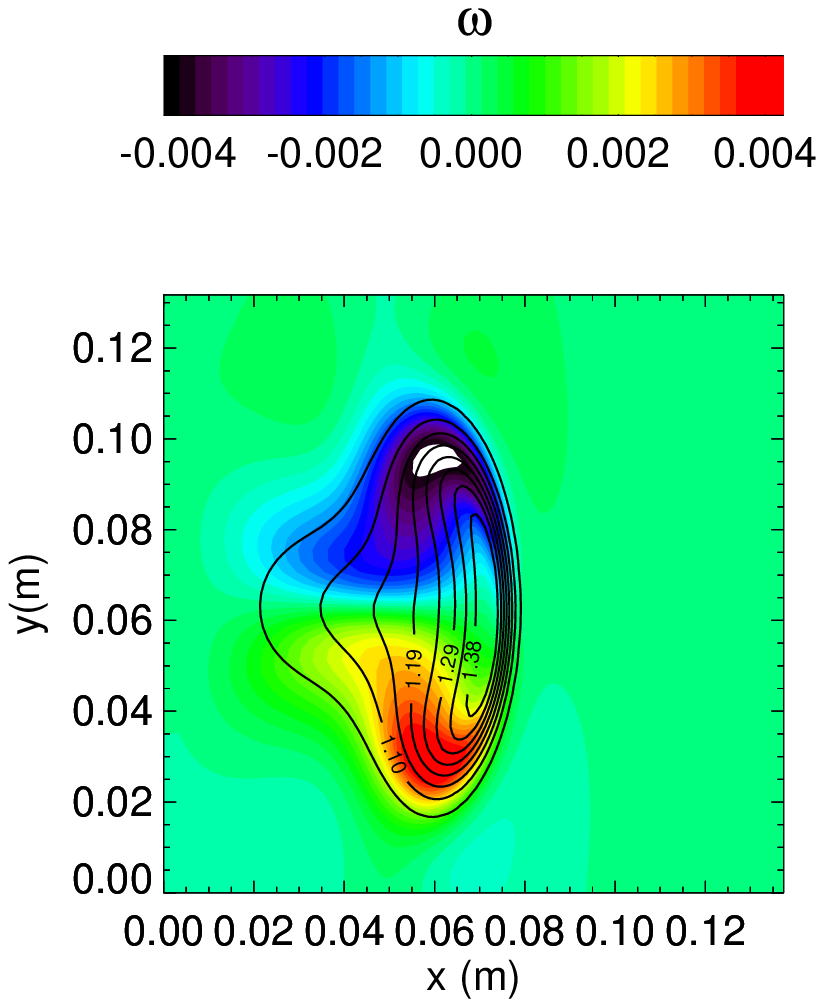}
    \label{morphology:int_vort}
    }
  \subfigure[]{
    \includegraphics[width=0.3\textwidth]{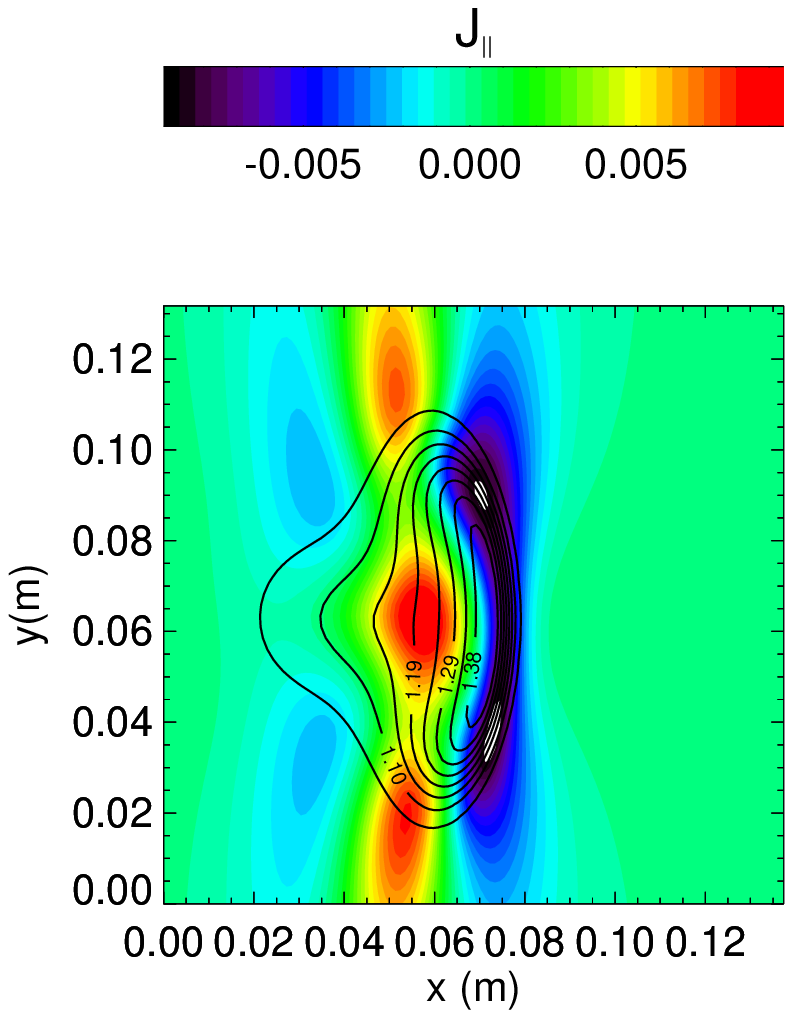}
    \label{morphology:int_jpar}
    }
  \subfigure[]{
    \includegraphics[width=0.3\textwidth]{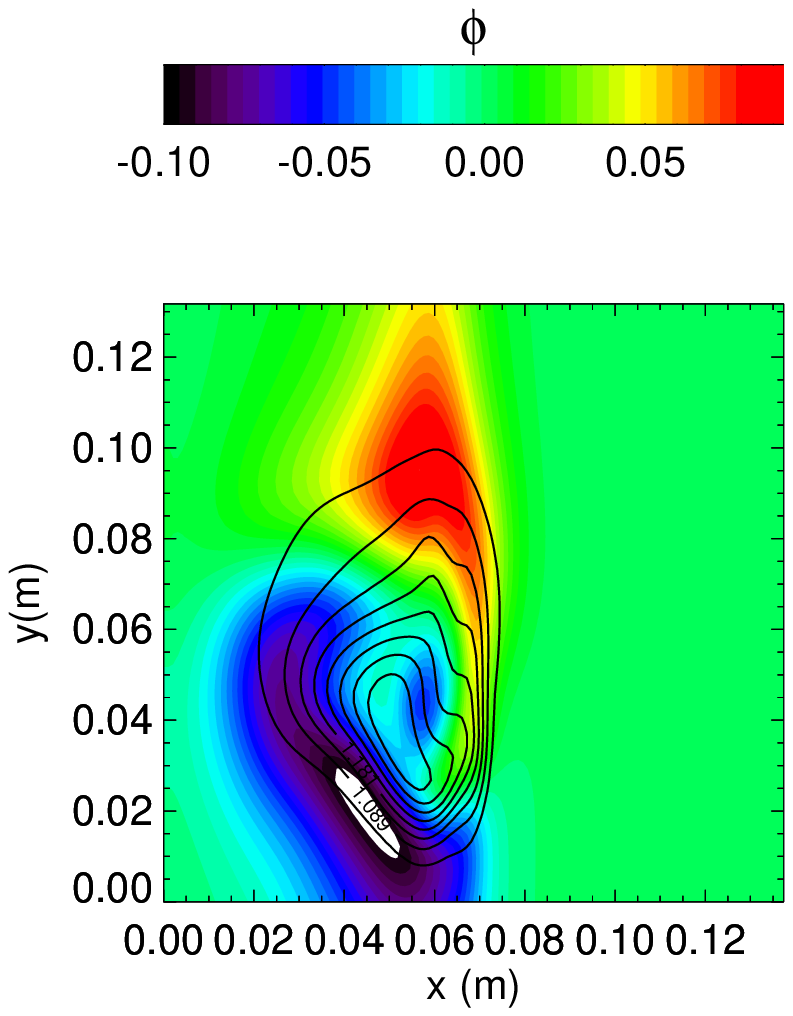}
    \label{morphology:Boltz_phi}
    }
  \subfigure[]{
    \includegraphics[width=0.3\textwidth]{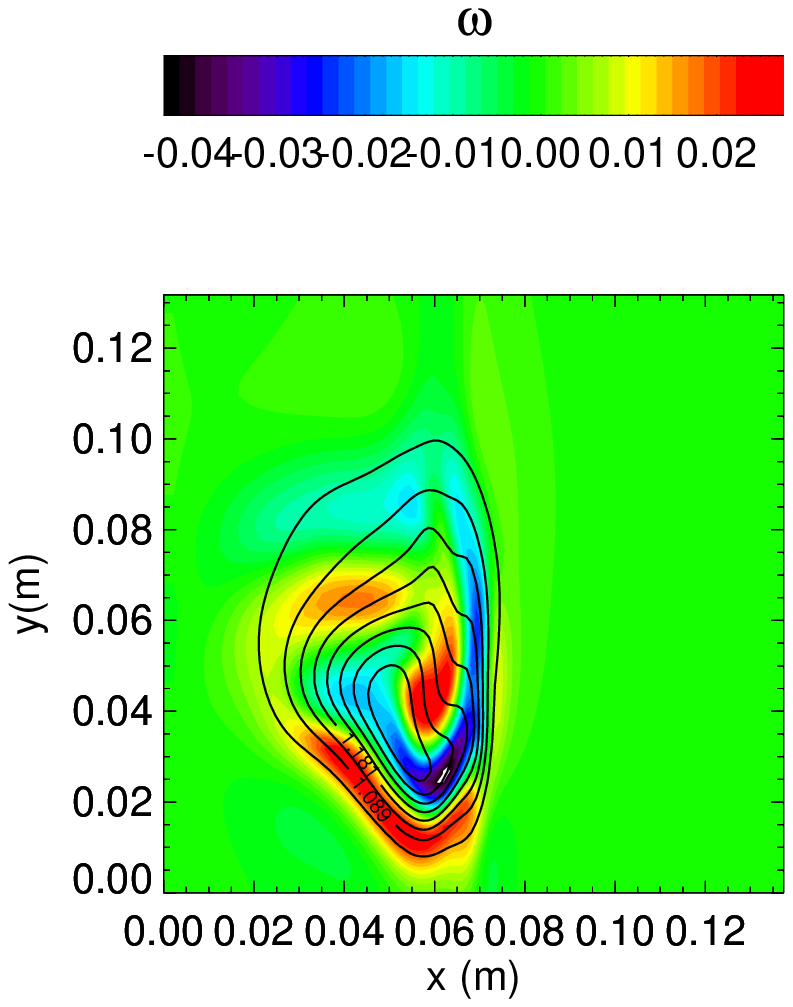}
    \label{morphology:Boltz_vort}
    }
  \subfigure[]{
    \includegraphics[width=0.3\textwidth]{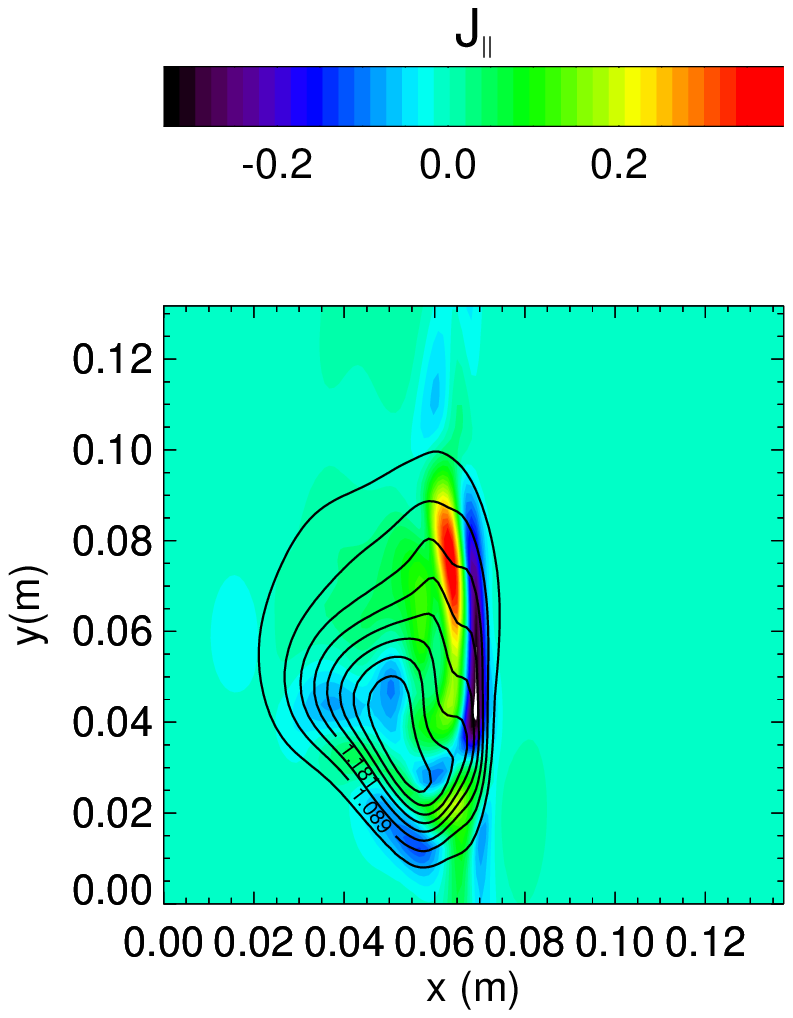}
    \label{morphology:Boltz_jpar}
    }
  \end{center}
\caption{Potential, vorticity and parallel current measured at the midplane of a filament in the interchange regime (a, b and c) and Boltzmann regime (d, e and f) with density overlaid as contour lines. Parallel current and Vorticity evolve on a smaller length scale in the Boltzmann regime, confirming the parallel current drive in the vorticity equation. Interchange data is sampled at $200\mu s$ and Boltzmann regime data is sampled at $100\mu 2$.}
\label{morphology}
\end{figure}

\begin{figure}[ht!]
\begin{center}
  \subfigure[]{
    \includegraphics[width=0.35\textwidth]{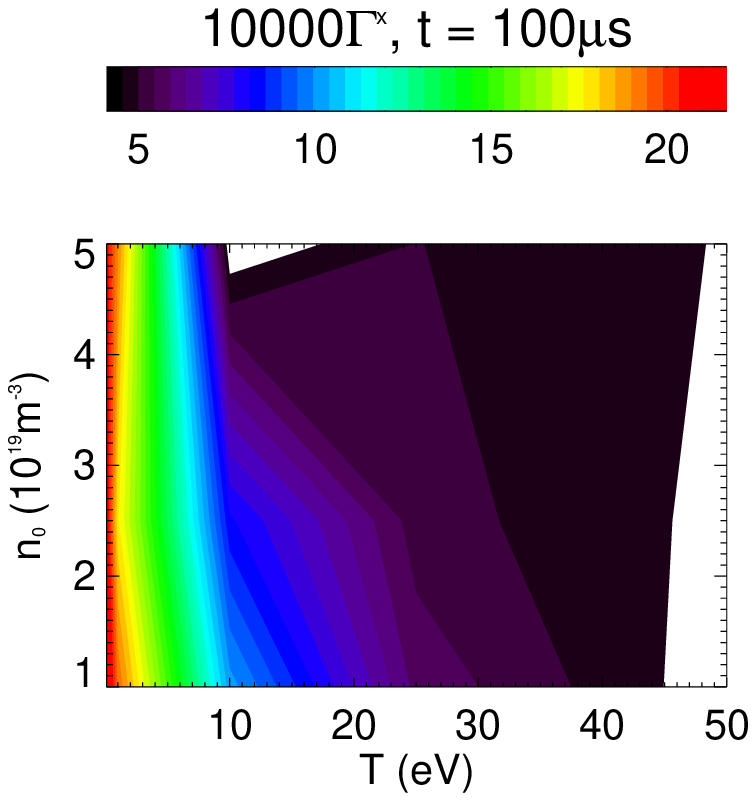}
    \label{trans_contour:x}
    }
   \subfigure[]{
    \includegraphics[width=0.35\textwidth]{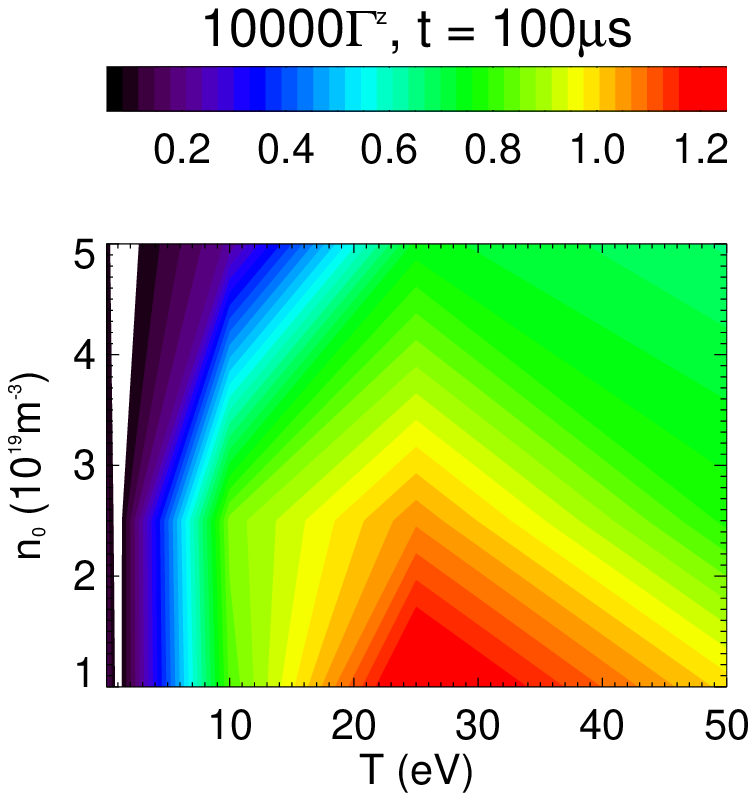}
    \label{trans_contour:z}
    }
   \subfigure[]{
    \includegraphics[width=0.35\textwidth]{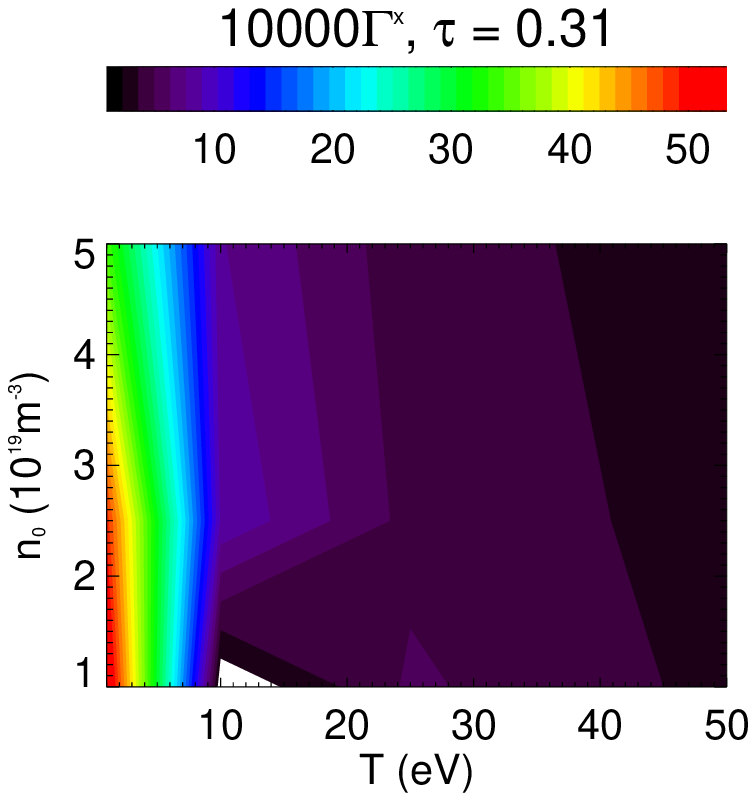}
    \label{trans_contour:x_cs}
    }
  \subfigure[]{
    \includegraphics[width=0.35\textwidth]{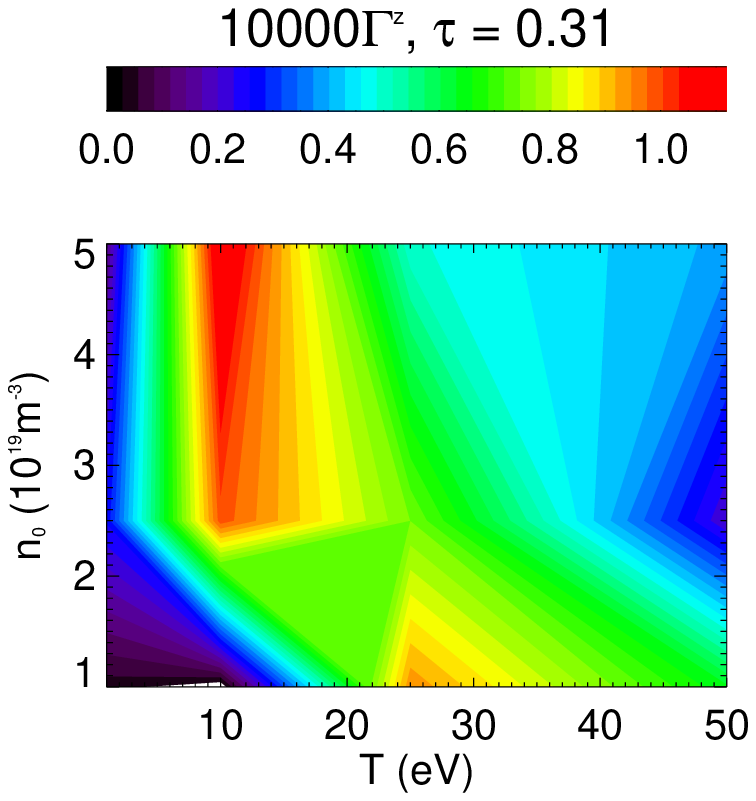}
    \label{trans_contour:z_cs}
    }
\end{center}
\caption{Volume averaged particle fluxes in $x$ and $z$ in the $n_{0},T_{e}$ parameter space sampled at $t = 40\mu s$ (upper) and $\tau = \Delta t c_{s}/L_{||} = 0.31$ (lower) . The contours are sparsely populated due to the excess computational time required to fill the contours more thoroughly, however underlying trends are still evident.}
\label{trans_contour}
\end{figure}

\begin{figure}[ht!]
  \begin{center}
    \subfigure[]{
      \includegraphics[width=0.45\textwidth]{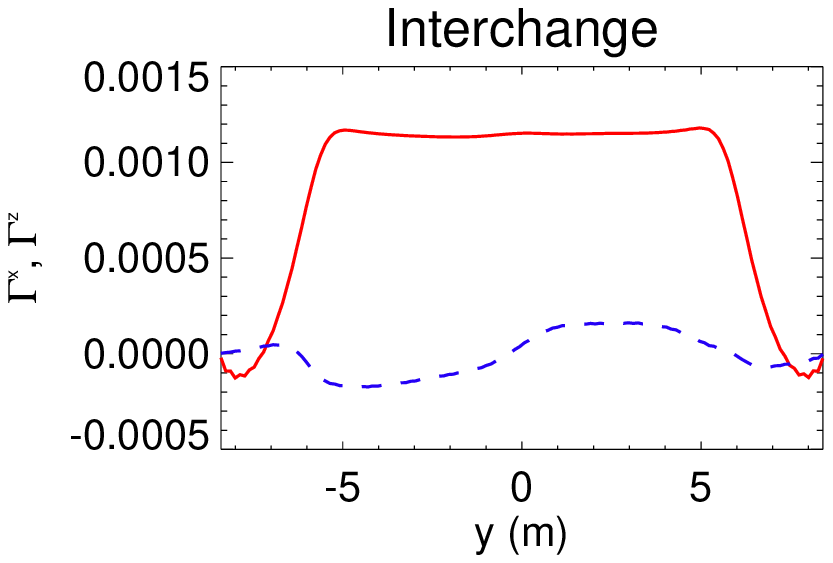}
      \label{trans_length:int}
      }
    \subfigure[]{
      \includegraphics[width=0.45\textwidth]{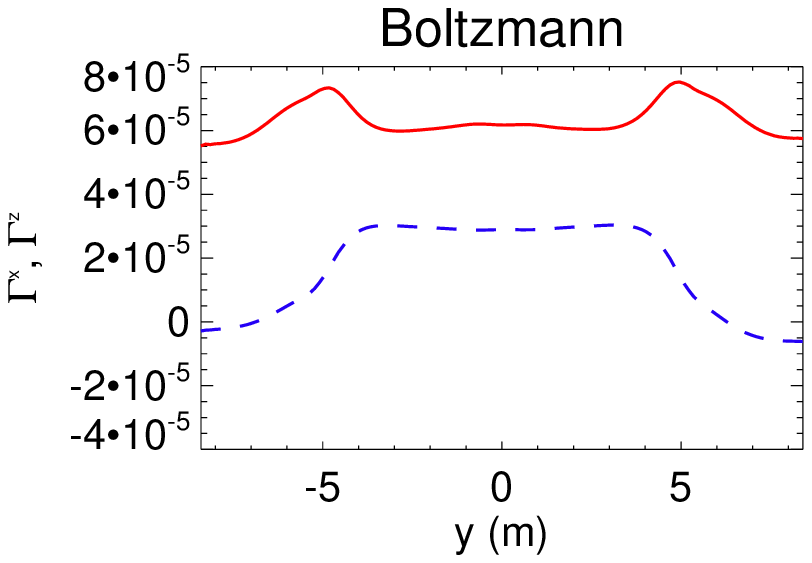}
      \label{trans_length:boltz}
      }
\end{center}
\caption{Parallel variation of $\Gamma_{x}$ (red, solid) and $\Gamma_{z}$ (blue, broken) averaged over the filament cross section in the Interchange regime (left) and the Boltzmann regime (right) at 1$ms$ and 100$\mu s$ respectively. }
\label{trans_length}
\end{figure}

\begin{figure}[ht!]
  \begin{center}
    \subfigure[]{
      \includegraphics[width=0.45\textwidth]{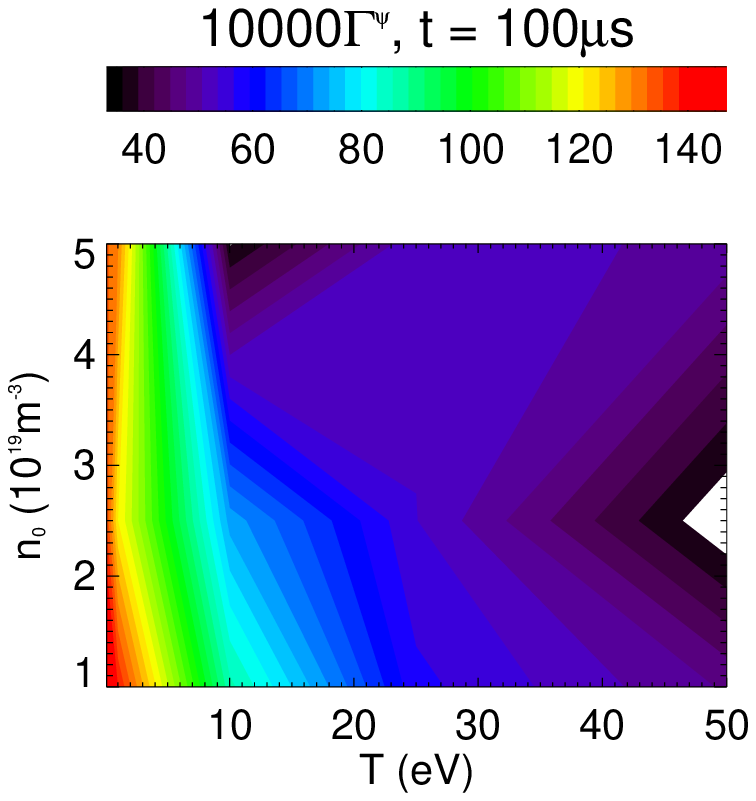}
      \label{trans_contour_psi:100mus}
      }
    \subfigure[]{
      \includegraphics[width=0.45\textwidth]{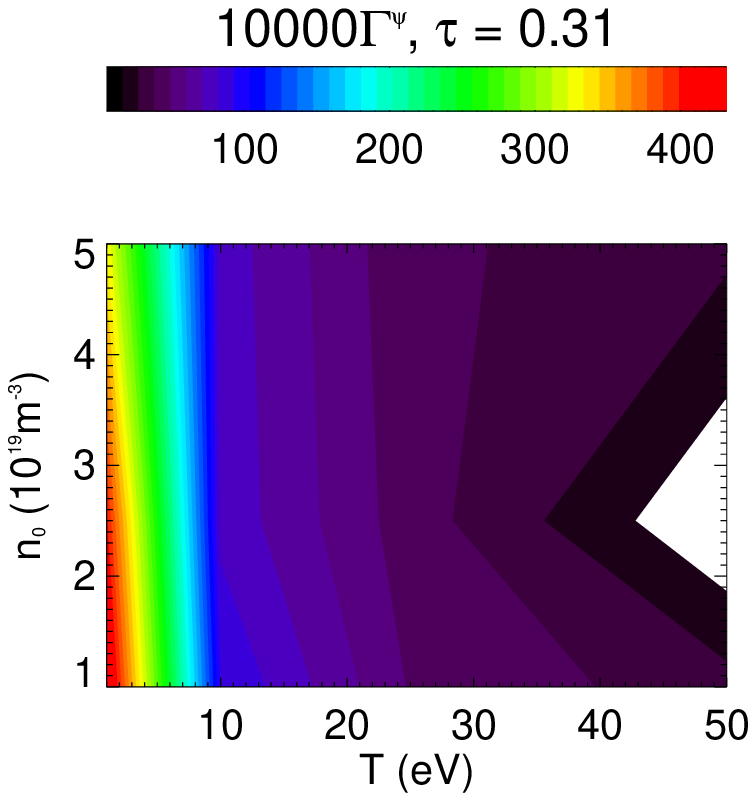}
      \label{trans_contour_psi:cs}
      }
    \end{center}
    \caption{Particle flux in the $\psi$ direction measured over the $n_{0}$ $T_{e}$ parameter space at $100\mu s$ (a) and $\tau = 0.31$ (b)}
    \label{trans_contour_psi}
\end{figure}

Simulations have been sampled at $100\mu s$ in \ref{corr_contour:abs_time} and at $\tau = 0.31$ in \ref{corr_contour:cs} where $\tau = c_{s}t/L_{||}$ is the time normalized to parallel streaming time. A clear transition occurs with temperature between a state with very low values of $C$ to a state with comparatively high values. The low $C$ state is the interchange regime where the symmetry of the dipolar potential structure ensures that $C\sim 0$ due to the volume average. When $C$ drops sharply below $0$ the system state is in the Boltzmann regime where the potential begins to overlap the density due to the Boltzmann response and a phase matching occurs. It is notable that a brief transitory period where $C$ grows positively occurs. It is likely that in this region both the interchange mechanism and the Boltzmann response are strongly affecting the filament. Figure \ref{int_xsec} presents a cross-field profile of the filament in the interchange regime which shows that the mushrooming structure associated with the interchange mechanism \cite{YuPoP2006,GarciaPoP2006,YuPoP2003} is reproduced when temperature is low.
Moving away from the midplane (figures \ref{int_xsec:lower} and \ref{int_xsec:upper}) the symmetry of the filament cross-section in the $x,z$ plane is broken. This is due to the enhanced component of $\xi^{\perp}$ by the magnetic shear, as described in section 4, which is antisymmetric along the length of the filament. As a result the lobe asymmetry develops anti symmetrically along the filament and therefore the volume average in (22) still leads to very small values of $C$ in this regime. 
\\ \\In the Boltzmann regime the cross-field evolution of the filament is markedly different. Figure \ref{Boltz_xsec} shows the cross-field evolution of the filament at the midplane. 
In the early stages of evolution only the interchange mechanism is driven. The Boltzmann response then acts to rotate the polarized charges, which reduces the advective electric field and eventually leads to the phase matching of the potential and density which gives rise to a significant correlation. Since charge conduction is much faster than in the interchange regime the polarization of the charge is weaker, which allows the dipole rotation to occur faster than the filament can be ejected outwards. This spins the filament. The filament straightens out (along the field line) as the potential acts to alleviate parallel density gradients, as shown in figure \ref{Boltz_par}. 
 In the final panel of figure \ref{Boltz_xsec} the filament has become highly non-linear and the advection term in the vorticity equation begins to dominate. This drives the filament towards a turbulent state where the advection non-linearity prevents the plasma from adopting a perfect Boltzmann response. The spatial morphology of the Boltzmann regime is shown in figure \ref{morphology} where it is also compared to the morphology in the interchange regime. 
Turbulent eddies appear as vorticity fluctuations in figure \ref{morphology:Boltz_vort} when the vorticity self-advection term becomes important. This is clearly the case in figures \ref{Boltz_xsec:3} and \ref{Boltz_xsec:4}. They occur on the same length scale as parallel current fluctuations whilst the density reacts on a larger length scale. This shows that it is not the density which drives the vorticity fluctuations (as is the case in the interchange regime) but rather the reaction to small scale current fluctuations. 
\\ \\Filaments contribute a large component of non-diffusive particle transport into the SOL. The level of this transport is determined by the turbulent conditions inside the seperatrix that leads to filament ejection and is beyond the scope of this paper. With filaments contributing such a significant proportion of SOL density, however, the subsequent redistribution of the density within the filament may play a crucial role in determining properties of the SOL; SOL width for example. In the 2D theory of blobs the transport of density within the blob is predominantly radial. It has been clearly demonstrated here that 2D blob theory based on interchange dynamics is only partly correct. Comparing the dynamics of the filament cross-section in the interchange and Boltzmann regimes shows that there is a significantly higher level of transport in the $x$ direction (which coincides with the radial direction at the midplane) in the interchange regime than in the Boltzmann regime. The transport of density can be quantified by a flux 
\begin{equation}
\Gamma_{x,z} =  v_{x,z}\delta n 
\end{equation}
 The transport is advective with the source of advection being the $\textbf{E}\times\textbf{B}$ velocity in the field aligned coordinate system defined by (15) which gives (see Appendix A for derivation)
\begin{equation}
\langle\frac{\Gamma_{x,z}}{n_{0}T_{e}}\rangle = \langle \delta n \frac{\partial \phi}{\partial z,x} \rangle
\end{equation}
Angled brackets once again represent a volume average. $\Gamma_{x}$ and $\Gamma_{z}$ have been calculated for the parameter space in figure \ref{corr_contour} with $n_{0} = \left[1,2.5,5\right]\times 10^{19}m^{-3}$ and $T_{e} = \left[0.1,1,10,25,50\right]eV$. The results are presented in figures \ref{trans_contour:x} and \ref{trans_contour:z} with all values calculated at $t=100\mu s$ and in \ref{trans_contour:x_cs} and \ref{trans_contour:z_cs} with all values calculated at $ 0.31\tau_{||}$. 
The results have been divided by $n_{0}T_{e}$ to eliminate the temperature and density dependency that these parameters introduce and focuses on the transport induced by the filament motion. In figures \ref{trans_contour:x} and \ref{trans_contour:x_cs} (upper panels) the expected behaviour of $\Gamma_{x}$ is observed; high levels of transport are observed in the interchange regime with a sharp reduction after the transition from interchange to Boltzmann dynamics. In the Boltzmann regime the fast conduction of charge halts the outward (in the $x$ direction) motion and the phase matching due to the Boltzmann response can spin the filament which acts to reduce radial transport further, but can enhance transport in the $z$ direction. This is not to say that transport in the $z$ direction is absent in the interchange regime. As demonstrated in figure \ref{int_xsec} the filament cross-section becomes asymmetric as one moves away from the midplane. Figure \ref{trans_length} shows the transport at different points along the filament by confining the average in (23) to the cross-section only. 
Significant transport in $z$ can occur in the interchange regime, however the symmetry around the midplane ensures that when volume averaged, the net transport is small. In the Boltzmann regime the symmetry around the midplane is broken and net transport in $z$ emerges, though the level of this transport at most points along the field line is drastically lower than in the interchange regime. The symmetry observed in the interchange regime is a result of the dependence of interchange dynamics on the magnetic parameters of the flux tube as demonstrated in section 4. Although the $x,y,z$ coordinate system is convenient for simulation purposes it is not particularly relevant to experiment. As such the particle flux in the $\psi$ direction has also been calculated and presented in figure \ref{trans_contour_psi}.
This is a quantity which can in principle be experimentally measured. The same trend as observed in figure \ref{trans_contour} is observed, with a sharp decrease in transport as the filament transitions to the from the interchange regime to the Boltzmann regime.

\section{Conclusion}
By employing a simulation geometry based on a flux tube in the SOL of a MAST L-mode DND plasma, parallel gradients have been shown to develop in filaments as a natural consequence of the magnetic geometry. These gradients are a result of variation in drive for the interchange mechanism along the length of the filament due to magnetic curvature, flux expansion and magnetic shear. The magnetic curvature effect is observed to be sub dominant to the latter two, which cause the filament to balloon at the midplane. Comparison with the two-region model of Myra \emph{et.al} \cite{MyraTwo-Region1} identifies the filament to be in the resistive ballooning regime. In this regime the filament motion is limited by inertia which allows filaments to obtain faster velocities than in the sheath limited model. This is observed in the simulations presented, where the dynamics of the filament between the two X-points are independent of the dynamics of the filament in the divertor region and of the boundary conditions at the divertor target. It is noted that this conclusion could be a good topic for an experimental investigation. In the simulations presented no cross-field variation in magnetic parameters were included. This prevents filaments from re-establishing an electrical connection to the sheath, as predicted in \cite{RyutovConPlasmaPhys2008}. Simulations including cross-field variation in the magnetic geometry are certainly a topic which should be pursued, especially with the development of more complicated divertor geometries such as the snowflake \cite{RyutovPoP2007,PirasPPCF2009,SoukhanovskiiNF2011,RyutovPPCF2012} or the Super-X \cite{ValanjuPoP2009,KatramadosFusEngDes2011,LisgoEPS2009} concepts. It is also important to recognise that no variation in the background parameters of the simulation along the field line have been included. In reality strong variation in both temperature and density are expected. To properly model the interaction between the filament and these background gradients a non-isothermal model is required. 
\\ \\The development of sustained parallel gradients in both density and potential can drive the filament away from the standard 2D interchange dynamics. At low collisionality these gradients force the electrostatic potential within the filament to adopt a Boltzmann response which can cause a monopolar rotation of the filament cross-section about its centre. At low collisionality fast charge conduction along the field line prevents a rapid build up of polarized charge and the spinning motion due to the Boltzmann response can occur faster than the ejective motion due to charge seperation. This drastically reduces radial transport, but can enhance transport in other directions.  By measuring the correlation between density and potential in $15$ simulations which span an $n_{0},T_{e}$ parameter space the transition between interchange and Boltzmann dynamics has been confirmed. The transition occurs as expected, with increasing temperature (and correspondingly decreasing collisionality). The approximations within the model, particularly the neglect of hot ions and the isothermal approximation, make the results only qualitatively accurate. The question of hot ions requires inclusion of finite Larmor orbit effects which will be addressed in future work. Furthermore the effects of electron inertia and electromagnetic effects have been neglected in the version of parallel Ohm's law used here. Electron inertia leads to a temporal delay in the response of the ions whilst a coupling to the magnetic field introduces Alfven waves into the system \cite{ScottPPCF97}. Both of these effects may be important in the seperatrix and near SOL regions \cite{MilitelloPPCF2011} and may alter the transition of the filament to the Boltzmann regime and are a good topic for future investigation.  The dynamic regime has a profound effect on the subsequent transport of density within the filament. In the interchange regime significantly higher levels of radial transport are observed, whilst the phase matching that occurs in the Boltzmann regime coupled with fast charge conduction halts the radial advection of the filament and suppresses radial transport. This suggests that only filaments governed by interchange dynamics at the seperatrix can reach the far SOL, with filaments in the Boltzmann regime confined to the region near the seperatrix. It is not clear if filaments born in one dynamic regime can transition to another dynamically. This will require inclusion of cross-field variation in either magnetic of background plasma parameters (or both). The idea of a filament in the Boltzmann regime cooling and transitioning to interchange dynamics is in line with experimental measurements of inter-ELM filaments on MAST \cite{AyedPPCF2009} where filaments are observed to propagate radially only after an initial stationary period. The model outlined in this paper suggests that such cooling is necessary in a wide range of operating parameters for MAST to facilitate the transport of filaments into the far SOL observed so frequently in experiments.

\section{Appendix A}
In this appendix a brief derivation of the $\textbf{E}\times\textbf{B}$ velocity in the field aligned coordinate system used to calculate tranport fluxes in (23) will be given. The contravariant and covariant metric tensors of the field aligned system (15) are 
\begin{equation}
g^{ij} \equiv \textbf{e}^i \cdot\textbf{e}^j =\left(\begin{array}{ccc}
\left(RB_{\theta}\right)^2 & 0 & -I\left(RB_{\theta}\right)^2 \\
0 & 1 / h_{\theta}^2 & -\nu / h_{\theta}^2 \\
-I\left(RB_{\theta}\right)^2 & -\nu / h_{\theta}^2 & I^2\left(RB_{\theta}\right)^2 + B^2 / \left(RB_{\theta}\right)^2 \end{array} \right)
\end{equation}
\begin{equation}
g_{ij} \equiv \textbf{e}_i \cdot\textbf{e}_j = \left(\begin{array}{ccc}
I^2 R^2 + 1 / \left(RB_{\theta}\right)^2 & B_{\zeta}h_{\theta} I R /B_{\theta}  & I R^2 \\
B_{\zeta}h_{\theta} I R /B_{\theta}  & B^2h_{\theta}^2 / B_{\theta}^2 & B_{\zeta} h_{\theta} R /B_{\theta}  \\
I R^2 & B_{\zeta}h_{\theta} R / B_{\theta} & R^2 \end{array} \right)
\end{equation}
In this system the magnetic field can be written
\begin{equation}
\textbf{B} = \nabla x \times \nabla z = \frac{1}{J}\textbf{e}_{y} = \frac{B_{\theta}}{h_{\theta}} \textbf{e}_{y}
\end{equation}
The magnetic field tangency vector $\textbf{b}$ can then be written 
\begin{equation}
\textbf{b} = \frac{1}{J B}\textbf{e}_{y} =  \frac{1}{J B}\left(g_{xy}\nabla x + g_{yy}\nabla y + g_{yz} \nabla_{z}\right)
\end{equation}
which gives the covariant components
\[ b_{x} = \frac{B_{\zeta}IR}{B} \]
\begin{equation}
 b_{y} = \frac{Bh_{\theta}}{B_{\theta}}
\end{equation}
\[ b_{z} = \frac{B_{\zeta}R}{B} \]
Since 
\begin{equation}
\nabla \phi = \frac{\partial \phi}{\partial x}\nabla x +  \frac{\partial \phi}{\partial y}\nabla y +  \frac{\partial \phi}{\partial z}\nabla z
\end{equation}
the contravariant components of $\textbf{b}\times\nabla\phi$ are 
\begin{equation} \left(\textbf{b}\times\nabla\phi\right)^{k} = \left(b_{i}\frac{\partial \phi}{\partial u^{j}} - b_{j}\frac{\partial \phi}{\partial u^{i}}\right)\nabla u^{i} \times \nabla u^{j}
= \frac{1}{J}\left(b_{i}\frac{\partial \phi}{\partial u^{j}} - b_{j}\frac{\partial \phi}{\partial u^{i}}\right)\textbf{e}_{k}
\end{equation}
Assuming that $\partial_{y}\phi << \partial_{x}\phi , \partial_{z}\phi$ 
\[\left(\textbf{b}\times\nabla\phi\right)^{x} = B\frac{\partial\phi}{\partial z}\]
\begin{equation}
\left(\textbf{b}\times\nabla\phi\right)^{y} = \frac{B_{\theta}B_{\phi}R}{h_{\theta}B}\frac{\partial\phi}{\partial x} - \frac{B_{\theta}B_{\phi}IR}{h_{\theta}B}\frac{\partial\phi}{\partial z}
\end{equation}
\[\left(\textbf{b}\times\nabla\phi\right)^{z} = -B\frac{\partial\phi}{\partial x}\]
Since the $\textbf{E}\times\textbf{B}$ velocity is given by $v_{E}^{j} = \left(\textbf{b}\times\nabla\phi\right)^{j}/B$ the flux of particles is given by 
\begin{equation}
\Gamma^{j} = \delta n v_{E}^{j}
\end{equation}
Finally normalizing the density to the background density, $n_{0}$ and the potential to the electron temperature (in $eV$) we arrive at 
\[ \frac{\Gamma^{x}}{n_{0}T_{e}} = \delta n \frac{\partial \phi}{\partial z} \]
\begin{equation}
  \frac{\Gamma^{z}}{n_{0}T_{e}} = \delta n \frac{\partial \phi}{\partial x}
\end{equation}
A more physically relevant flux to calculate is $\Gamma^{\psi}$ since this can be measured experimentally. To calculate $\Gamma^{\psi}$ the derived form of $\textbf{b}\times\nabla \phi$ is projected onto $\textbf{e}_{\psi}$ to give
\begin{equation}
\Gamma^{\psi} = \Gamma \cdot \textbf{e}_{\psi} = \frac{\Gamma^{x}}{RB_{\theta}}
\end{equation}
\section{References}
\bibliographystyle{prsty}
\bibliography{Paper-ref}

\begin{thebibliography}{10}

\bibitem{BoedoPoP2001}
J.~A. Boedo {\it et~al.}, Phys. Plasmas {\bf 8},  11  (2001).

\bibitem{KirkPPCF2006}
A. Kirk {\it et~al.}, Plasma Phys. Control. Fusion {\bf 48},  B433  (2006).

\bibitem{DudsonPPCF2005}
B.~D. Dudson {\it et~al.}, Plasma Phys. Control. Fusion {\bf 47},  885  (2005).

\bibitem{DudsonPPCF2008}
B.~D. Dudson {\it et~al.}, Plasma Phys. Control. Fusion {\bf 50},  124012
  (2008).

\bibitem{CarrerasPoP2001}
B.~A. Carreras, V.~E. Lynch, and B. LaBombard, Phys. Plasmas {\bf 8},  3702
  (2001).

\bibitem{EndlerNF95}
M. Endler {\it et~al.}, Nucl. Fusion {\bf 35},  1307  (1995).

\bibitem{ZwebenPoP2002}
S.~J. Zweben {\it et~al.}, Phys. Plasmas {\bf 9},  1981  (2002).

\bibitem{FasoliPoP2006}
A. Fasoli {\it et~al.}, Phys. Plasmas {\bf 13},  055902  (2006).

\bibitem{AntarPRL2001}
G.~Y. Antar {\it et~al.}, Phys. Rev. Lett. {\bf 87},  065001  (2001).

\bibitem{YamadaNATURE2008}
T. Yamada {\it et~al.}, Nature Lett. {\bf 4},  721  (2008).

\bibitem{AyedPPCF2009}
N.~B. Ayed {\it et~al.}, Plasma Phys. Control. Fusion {\bf 51},  035016
  (2009).

\bibitem{KrasheninnikovPhysLet2001}
S.~I. Krasheninnikov, Phys. Lett. A {\bf 283},  368  (2001).

\bibitem{NedospasovNF85}
A.~D. Nedospasov, V.~G. Petrov, and G.~N. Fidel'man, Nucl. Fusion {\bf 25},  21
   (1985).

\bibitem{KrasheninnikovCzechJourn2005}
S.~I. Krasheninnikov, A.~I. Smolyakov, G.~Q. Yu, and T.~K. Soboleva, Czech.
  Journ. Phys. {\bf 55},  307  (2005).

\bibitem{YuPoP2006}
G.~Q. Yu, S.~I. Krasheninnikov, and P.~N. Guzdar, Phys. Plasmas {\bf 13},
  042508  (2006).

\bibitem{GarciaPoP2006}
O.~E. Garcia, N.~H. Bian, and W. Fundamenski, Phys. Plasmas {\bf 13},  082309
  (2006).

\bibitem{BianPoP2003}
N.~H. Bian, S. Bendekka, J.~V. Paulsen, and O.~E. Garcia, Phys. Plasmas {\bf
  10},  3  (2003).

\bibitem{MyraPoP2004}
J.~R. Myra, D.~A. D'Ippolito, S.~I. Krasheninnikov, and G.~W. Yu, Phys. Plasmas
  {\bf 11},  9  (2004).

\bibitem{YuPoP2003}
G.~Q. Yu and S.~I. Krasheninnikov, Phys. Plasmas {\bf 10},  4413  (2003).

\bibitem{JovanovicPoP2008}
D. Jovanovic, P.~K. Shukla, and F. Pegoraro, Phys. Plasmas {\bf 15},  112305
  (2008).

\bibitem{RozhanskyPPCF2008}
V. Rozhansky and A. Kirk, Plasma Phys. Control. Fusion {\bf 50},  025008
  (2008).

\bibitem{MadsenPoP2011}
J. Madsen {\it et~al.}, Phys. Plasmas {\bf 18},  112504  (2011).

\bibitem{KrasheninnikovPoP2003}
S.~I. Krasheninnikov and A.~I. Smolyakov, Phys. Plasmas {\bf 10},  3020
  (2003).

\bibitem{D'IppolitoPoP2004}
D.~A. D'Ippolito, J.~R. Myra, D.~A. Russell, and G.~Q. Yu, Phys. Plasmas {\bf
  11},  4603  (2004).

\bibitem{D'IppolitoContribPlasPhys2004}
D.~A. D'Ippolito {\it et~al.}, Contrib. Plasma Phys. {\bf 44},  205  (2004).

\bibitem{RussellPRL2004}
D.~A. Russell {\it et~al.}, Phys. Rev. Lett. {\bf 93},  265001  (2004).

\bibitem{BodiPoP2008}
K. Bodi, S.~I. Krasheninnikov, and A.~I. Smolyakov, Phys. Plasmas {\bf 15},
  102304  (2008).

\bibitem{D'IppolitoReview}
D.~A. D'Ippolito, J.~R. Myra, and S.~J. Zweben, Phys. Plasmas {\bf 18},  060501
   (2011).

\bibitem{KatzPRL2008}
N. Katz {\it et~al.}, Phys. Rev. Lett. {\bf 101},  015003  (2008).

\bibitem{AngusPRL2012}
J.~R. Angus, M.~V. Umansky, and S.~I. Krasheninnikov, Phys. Rev. Lett. {\bf
  108},  215002  (2012).

\bibitem{AngusPoP2012}
J.~R. Angus, S.~I. Krasheninnikov, and M.~V. Umansky, Phys. Plasmas {\bf 19},
  082312  (2012).

\bibitem{BOUT++}
B.~D. Dudson {\it et~al.}, Comp. Phys. Comm. {\bf 180},  1467  (2009).

\bibitem{EFIT}
L.~L. Lao {\it et~al.}, Nucl. Fusion {\bf 25},  1611  (1985).

\bibitem{Braginskii}
S.~I. Braginskii, \emph{Reviews of Modern Physics} {\bf 1},  205  (Consultants
  Bureau, New York, 1965).

\bibitem{SimakovPoP2003}
A.~N. Simakov and P.~J. Catto, Phys. Plasmas {\bf 20},  4744  (2003).

\bibitem{SimakovPoP2004}
A.~N. Simakov and P.~J. Catto, Phys. Plasmas {\bf 11},  2326  (2004).

\bibitem{KirkPPCF2004}
A. Kirk {\it et~al.}, Plasma Phys. Control. Fusion {\bf 46},  1591  (2004).

\bibitem{KocanPPCF2010}
M. Kocan and J.~P. Gunn, Plasma Phys. Control. Fusion {\bf 52},  045010
  (2010).

\bibitem{KocanJNucMat2011}
M. Kocan {\it et~al.}, J. Nucl. Mater. {\bf 415},  S1133  (2011).

\bibitem{TamainJNucMat2011}
P.~A. Tamain {\it et~al.}, J. Nucl. Mater. {\bf 415},  S1139  (2011).

\bibitem{ElmorePPCF2012}
S. Elmore {\it et~al.}, Plasma Phys. Control. Fusion {\bf 54},  065001  (2012).

\bibitem{ElmoreFusEngDes2013}
S. Elmore {\it et~al.}, Fus. Eng. Des.  10.1016/j.jnucmat.2013.01.268  (2013).

\bibitem{AllanJNucMat2013}
S.~Y. Allan {\it et~al.}, J. Nucl. Mater.
  http://dx.doi.org/10.1016/j.jnucmat.2013.01.263  (2013).

\bibitem{BisaiPoP2012}
N. Bisai, R. Singh, and P.~K. Kaw, Phys. Plasmas {\bf 19},  052509  (2012).

\bibitem{AngusDriftWave2012}
J.~R. Angus and S.~I. Krasheninnikov, Phys. Plasmas {\bf 19},  052504  (2012).

\bibitem{MilitelloPPCF2013}
F. Militello {\it et~al.}, Plasma Phys. Control. Fusion {\bf 55},  025005
  (2013).

\bibitem{GarciaPPCF2006}
O.~E. Garcia {\it et~al.}, Plasma Phys. Control. Fusion {\bf 48},  L1  (2006).

\bibitem{XuBOUT}
X.~Q. Xu, M.~V. Umansky, B.~D. Dudson, and P.~B. Snyder, Comm. Comp. Phys. {\bf
  4},  949  (2008).

\bibitem{FarinaNF1993}
D. Farina, R. Pozzoli, and D.~D. Ryutov, Nucl. Fusion {\bf 13},  1315  (1993).

\bibitem{RyutovConPlasmaPhys2008}
D.~D. Ryutov and R.~H. Cohen, Contrib. Plasma Phys. {\bf 48},  48  (2008).

\bibitem{RyutovConPlasmaPhys2006}
D.~D. Ryutov and R.~H. Cohen, Contrib. Plamsa Phys. {\bf 46},  678  (2006).

\bibitem{MyraTwo-Region1}
J. Myra, D.~A. Russell, and D.~A. D'Ippolito, Phys. Plasmas {\bf 13},  112502
  (2006).

\bibitem{RyutovPoP2007}
D.~D. Ryutov, Phys. Plasmas {\bf 14},  064502  (2007).

\bibitem{PirasPPCF2009}
F. Piras {\it et~al.}, Plasma Phys. Control. Fusion {\bf 51},  055009  (2009).

\bibitem{SoukhanovskiiNF2011}
V.~A. Soukhanovskii {\it et~al.}, Nucl. Fusion {\bf 51},  012001  (2011).

\bibitem{RyutovPPCF2012}
D.~D. Ryutov, R.~H. Cohen, T.~D. Rognlien, and M.~V. Umansky, Plasma Phys.
  Control. Fusion {\bf 54},  124050  (2012).

\bibitem{ValanjuPoP2009}
P.~M. Valanju, M. Kotschenreuther, S.~M. Mahajan, and J. Canik, Phys. Plasmas
  {\bf 16},  056110  (2009).

\bibitem{KatramadosFusEngDes2011}
I. Katramados {\it et~al.}, Fusion Eng. Des  (2011).

\bibitem{LisgoEPS2009}
S. Lisgo {\it et~al.}, 36th EPS Conference on Plasma Phys. {\bf 33E},  0
  (2009).

\bibitem{ScottPPCF97}
B. Scott, Plasma Phys. Control. Fusion {\bf 39},  1635  (1997).

\bibitem{MilitelloPPCF2011}
F. Millitello and W. Fundamenski, Plasma Phys. Control. Fusion {\bf 53},
  095002  (2011).

\end{thebibliography}

\section{Acknowledgements}
We thank F. Militello for his extensive help in preparing this paper. We also thank A. Kirk and W. Morris for a number of discussions which have helped to shape the work.
\\Some simulations made use of resources on the HECToR super-computer, provided through EPSRC grant EP/H002081/1. 
\\This work was part-funded by the RCUK Energy Programme [grant number EP/I501045] and the European Communities under the contract of Association between EURATOM and CCFE.  To obtain further information on the data and models underlying this paper please contact PublicationsManager@ccfe.ac.uk.  The views and opinions expressed herein do not necessarily reflect those of the European Commission.

\end{document}